\documentclass[final,10pt,journal,compsoc]{IEEEtran}

\ifCLASSOPTIONcompsoc
  \usepackage[nocompress]{cite}
\else
  \usepackage{cite}
\fi

\usepackage{microtype}
\usepackage{graphicx}
\usepackage[caption=false]{subfig}  
\usepackage{booktabs} 
\usepackage{amsmath} 
\newcommand{\plotwidth}{.285\textwidth}  

\newcommand{\eat}[1]{}

\usepackage{ifdraft}
\ifoptionfinal{
        
        \newcommand{\added}[1]{#1}
        \newcommand{\deleted}{\eat}
        \newcommand{\replaced}[2]{#1}
}{
        \usepackage{changes}  
}

\newcommand{\twofig}[4]{
    \centering
    \subfloat[#1]{\makebox[.5\textwidth]{\includegraphics[width=\plotwidth]{PLOTS/#2.pdf}}}
    \subfloat[#3]{\makebox[.5\textwidth]{\includegraphics[width=\plotwidth]{PLOTS/#4.pdf}}}
}

\newcommand{\fourfig}[8]{
    \twofig{#1}{#2}{#3}{#4} \\
    \twofig{#5}{#6}{#7}{#8}
}

\usepackage{hyperref}

\usepackage[capitalise,noabbrev]{cleveref}
\usepackage{multirow}
\usepackage{balance}

\usepackage{tikz}
\usepackage{textcomp}
\usepackage{hyperref}
\usepackage{lipsum}

\newcommand\copyrighttext{%
  \footnotesize \textcopyright 2022 IEEE. Personal use of this material is permitted.
  Permission from IEEE must be obtained for all other uses, in any current or future
  media, including reprinting/republishing this material for advertising or promotional
  purposes, creating new collective works, for resale or redistribution to servers or
  lists, or reuse of any copyrighted component of this work in other works.
  DOI: \href{https://doi.org/10.1109/TPDS.2022.3146195}{10.1109/TPDS.2022.3146195}}
\newcommand\copyrightnotice{%
\begin{tikzpicture}[remember picture,overlay]
\node[anchor=south,yshift=10pt] at (current page.south) {\fbox{\parbox{\dimexpr\textwidth-\fboxsep-\fboxrule\relax}{\copyrighttext}}};
\end{tikzpicture}%
}

\begin{document}


\title{The Supermarket Model with \\ Known and Predicted Service Times}

\author{Michael Mitzenmacher and Matteo Dell'Amico%
\IEEEcompsocitemizethanks{
\IEEEcompsocthanksitem M. Mitzenmacher is with the School of Engineering and Applied Sciences, Harvard University.  Email: {michaelm@eecs.harvard.edu}.  This work was supported in part by NSF grants CCF-2101140, DMS-2023528, CCF-1563710, and CCF-1535795.
\IEEEcompsocthanksitem M. Dell'Amico is with the University of Genoa. This work was carried out while he was with NortonLifeLock Research Group and EURECOM, France. Email: {della@linux.it}.
}}

\IEEEtitleabstractindextext{%
\begin{abstract}
The supermarket model refers to a system with a large number
of queues, where new customers choose $d$
queues at random and join the one with the fewest customers.  This
model demonstrates the power of even small amounts of
choice, as compared to simply joining a queue chosen uniformly at
random, for load balancing systems.  In this work we perform
simulation-based studies to consider variations where service times
for a customer are {\em predicted}, as might be done in modern
settings using machine learning techniques or related mechanisms.
Our primary takeaway is that using even seemingly weak
predictions of service times can yield significant benefits over blind
First In First Out queueing in this context.  However, some care must
be taken when using predicted service time information to both choose
a queue and order elements for service within a queue; while in many
cases using the information for both choosing and ordering is
beneficial, in many of our simulation settings we find that simply
using the number of jobs to choose a queue is better when using
predicted service times to order jobs in a queue.
In our simulations, we evaluate both synthetic and real-world
workloads--in the latter, service times are predicted by machine learning.
Our results provide practical guidance for the design of real-world systems;
moreover, we leave many natural theoretical open questions for future work,
validating their relevance to real-world situations.
\end{abstract}

\begin{IEEEkeywords}  
Supermarket model, prediction methods, scheduling, queueing analysis
\end{IEEEkeywords}
}

\maketitle
\copyrightnotice

\IEEEdisplaynontitleabstractindextext

\section{Introduction}
The success of machine learning (ML) has opened up new opportunities in
terms of improving the efficiency of a wide array of processes.  In
this paper, we consider opportunities for using ML
predictions in a specific setting: queueing in large distributed
systems using \replaced{the supermarket model}{``the power of two choices''}.  This also leads us to
consider variants of these systems that appear to have not previously
studied that do not use predictions as a starting point.

Scheduling with predicted service times is a topic of obvious
practical importance for system design; however, there is a relative
scarcity of studies exploring \replaced{the area}{the field}. We believe that this may be
because the community tends to prefer analytical studies,
and this particular topic appears particularly challenging to tackle
analytically, even when simplifying assumptions to facilitate
modeling are taken.

We take a different approach, basing our study on simulations. This
choice allows us to obtain new insight that is relevant to practical
system design issues. Moreover, besides using simulations based on standard distributions for queueing studies, we use real-world traces for system
evaluation, where job submission time, service time, \emph{and
predicted service time} come from real-world systems: in
particular, service time predictions are the output of a
state-of-the-art approach~\cite{amvrosiadis2018diversity}.


We start with \added{the} key background.  In queueing settings, the supermarket
model (also described as the power of two choices, or balanced
allocations) is typically described in the following way.  Suppose we
have a system of $n$ First In, First Out (FIFO) queues.
Jobs\footnote{In this paper we use jobs instead of the more specific
term customers, as the model applies to a variety of load-balancing
settings.} arrive to the system as a Poisson process of rate $\lambda
n$, and service times are independent and exponentially distributed
with mean 1.  If each job selects a random queue on arrival, then via
Poisson splitting \cite[Section 8.4.2]{DBLP:books/daglib/0012859} each
queue acts as a standard M/M/1 queue, and in equilibrium the fraction
of queues with {\em at least} $i$ jobs is $\lambda^i$.  Note that we
consider here the tails of the queue length distribution, as it makes
for easier comparisons.  If each job selects \deleted{a} two random queues on
arrival, and chooses to wait at the queue with fewer customers
(breaking ties randomly), then in the limiting system as $n$ grows to
infinity, in equilibrium the fraction of queues with at least $i$ jobs
is $\lambda^{2^i-1}$.  That is, the tails decrease doubly
exponentially in $i$, instead of single exponentially.  In practice,
even for moderate values of $n$ (say in the large hundreds), one
obtains performance close to this mean field limit; this can
\deleted{be} be proven based on appropriate concentration bounds.  More
generally, for $d$ choices where $d$ is an integer constant greater
than 1, the fraction of queues with at least $i$ jobs falls like
$\lambda^{(d^i-1)/(d-1)}$ \cite{DBLP:journals/tpds/Mitzenmacher01,
vvedenskaya1996queueing}.  While there are many variations on
the supermarket model and its analysis (see, e.g., \cite{aghajani2015mean,bramson2010randomized,mitzenmacher1999asymptotics,vocking2003asymmetry}), here we focus on this
standard, simple formulation, although we allow for general service distributions instead of just the exponential distribution.

As stated previously, here we study variations of the supermarket model where
service times are predicted.  To describe our work and
goals, we start by considering the baseline where service times are
known.

The analysis for the basic model described above
assumes that service times are exponentially distributed but specific job service times are not known. (Extensions of the analysis to more general distributions are known \cite{aghajani2015mean,bramson2010randomized}.)
As such, an incoming job uses only the number of jobs at each chosen
queue to decide which queue to join.  As both a theoretical question and for
possible practical implementations, it seems worthwhile to know what
further improvement is possible if service times of the jobs were
known.  

Recently, Hellemans and Van Houdt proved results in the
supermarket model setting where {\em job reservations} are made at $d$
randomly chosen queues, and once the first reservation reaches the
point of obtaining service, the other reservations are canceled.
This corresponds to choosing the least loaded (in terms of total
remaining service time\footnote{We use service time and
processing time interchang\added{e}ably\deleted{ in this paper}; both terms have been used
historically.}) of $d$ queues using FIFO queues.  Their work applies to
general service distributions; for the class of phase-type service
distributions, they are able to express the limiting behavior of the
system in terms of delayed differential
equations \cite{DBLP:journals/pomacs/HellemansH18}.  Their 
results, \replaced{including both theorems and simulations}{including theorems regarding the system behavior as well
as simulations}, show that using service time information can lead to significant
improvements in the average time a job spends in the system.  (The subsequent
work \cite{DBLP:journals/pomacs/HellemansBH19} examines several additional variations.)
Because of space limitations, we leave discussion
of the challenges of extending these results beyond FIFO queues to
\cref{app:cq1}.

However, when the service times are known, there are two possible ways
to potentially improve performance.  First, as above, one can use the service
times when selecting a queue, by choosing the least loaded queue.
Second, one can order the jobs using a strategy other than FIFO; the
natural strategies to minimize the average time in the system
(response time) are {\em shortest job first} (SJF), {\em preemptive
shortest job first} (PSJF), and {\em shortest remaining processing
time} (SRPT).  Here shortest job first assumes no preemption and
always schedules the job with the smallest service time when a job
completes, preemptive shortest job first allows preemption so that a
job with a smaller service time can preempt the running job, and
shortest remaining processing time allows preemption but is based on
the remaining processing time instead of the total service time for a
job.  Note that here we assume a preempted job does not need to start
from the beginning and can later continue service where it left off.
Also, while apparently somewhat less natural, PSJF allows
job priorities to be assigned on arrival to a queue without the need
for updating, unlike SRPT.

In the setting of a single queue, Mitzenmacher has recently considered
the setting where service time \replaced{is}{are} predicted rather than known
exactly \cite{mitz2019scheduling}.  In this model, the jobs have a
joint service-predicted service density function $g(x,y)$, where $x$
is the true service time and $y$ is the predicted service time.  He
provides formulae for the average \replaced{time in system}{response time} using
corresponding strategies {\em shortest predicted job first} (SPJF),
{\em preemptive shortest predicted job first} (PSPJF), and {\em
shortest predicted remaining processing time} (SPRPT).  Simulation
results suggest that in the single queue setting even weak predictors
can greatly improve performance over FIFO queues.  However, using the
\replaced{supermarket model}{power of two choices} already provides great improvements in systems
with multiple queues.  It is therefore natural to consider whether predictions
would still provide significant gains in the supermarket
model.


The contributions of this paper include the following.
\begin{itemize}
\item For the case of known service times, we provide a simulation study with synthetic traces
showing the potential gains when using SJF, PSJF, and SRPT queues in the supermarket model, providing an appropriate baseline. 
\item We similarly through simulations examine the benefits 
when only predicted information is available, using FIFO, SPJF, PSPJF, and SPRPT queues.  Here we use both synthetic and real-world datasets.  
\item We determine somewhat counterintuitive behaviors;  for example, we find
many cases where choosing the predicted least loaded queue performs worse than simply choosing the shortest queue.
\item We provide a number of open questions related to the analysis and
use of these systems.  
\end{itemize}


\section{Additional Related Work}
\label{sec:related}

The \replaced{supermarket model}{power of two choices} was first analyzed in the discrete settings
of hashing, modeled as balls and bins processes \cite{DBLP:journals/siamcomp/AzarBKU99,DBLP:journals/algorithmica/KarpLH96,DBLP:journals/cpc/Mitzenmacher99}.  It
was subsequently analyzed in the setting of queueing systems, in
particular in the mean field limit (also referred to as the fluid
limit) as the number of queues grows to infinity \cite{DBLP:journals/tpds/Mitzenmacher01,vvedenskaya1996queueing}.

Ordering jobs by service time has been studied extensively in single
queues.  The text \cite{harchol2013performance} provides an excellent
introduction to the analysis of standard approaches such as SJF and
\replaced{SRPT}{SPRT} in the single queue setting.  

Our work falls into a recent line of work that aims to use machine
learning predictions to improve traditional algorithms.  For example,
Lykouris and Vassilvitskii \cite{DBLP:conf/icml/LykourisV18} show how
to use prediction advice from ML algorithms to improve
online algorithms for caching in a way that provides provable
performance guarantees, using the framework of competitive analysis.
Other recent works with this theme include the development of learned Bloom filters
\cite{TCFLIS,mitzenmacher2018model} and heavy hitter algorithms that use
predictions \cite{hsu2018learning}.  One prior work in this vein has 
specifically
looked at scheduling with predictions in the setting of a fixed collection of
jobs, and consider variants of shortest predicted processing time that 
yield good performance in terms of the competitive ratio, with the performance
depending on the accuracy of the predictions \cite{purohit2018improving}. 

In scheduling of queues, some works have looked at the effects of using
imprecise information, including for load balancing in multiple queue
settings.  For example, Mitzenmacher considers using old load
information to place jobs (in the context of the \replaced{supermarket model}{power of two choices})
\cite{DBLP:journals/tpds/Mitzenmacher00}. A strategy called TAGS
studies an approach to utilizing multiple queues when no information
exists about the service time; jobs that run more than some threshold
in the first queue are canceled and passed to the second queue, and
so on \cite{DBLP:journals/jacm/Harchol-Balter02}.  

\begin{figure*}[htbp]
    \fourfig
    {Exponential service times, queue chosen by least loaded}{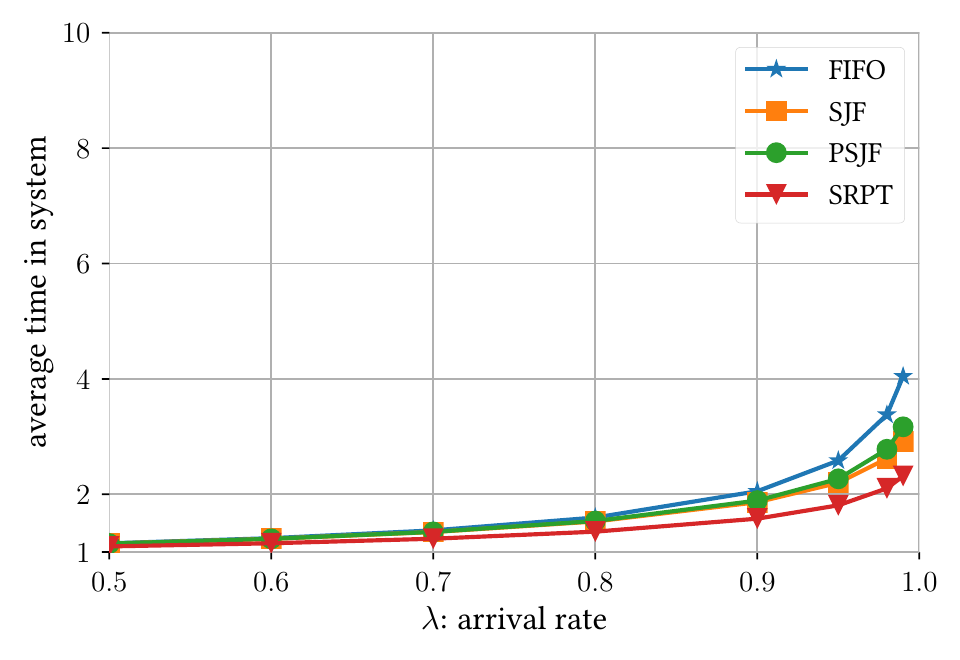}
    {Exponential service times, queue chosen by shortest queue}{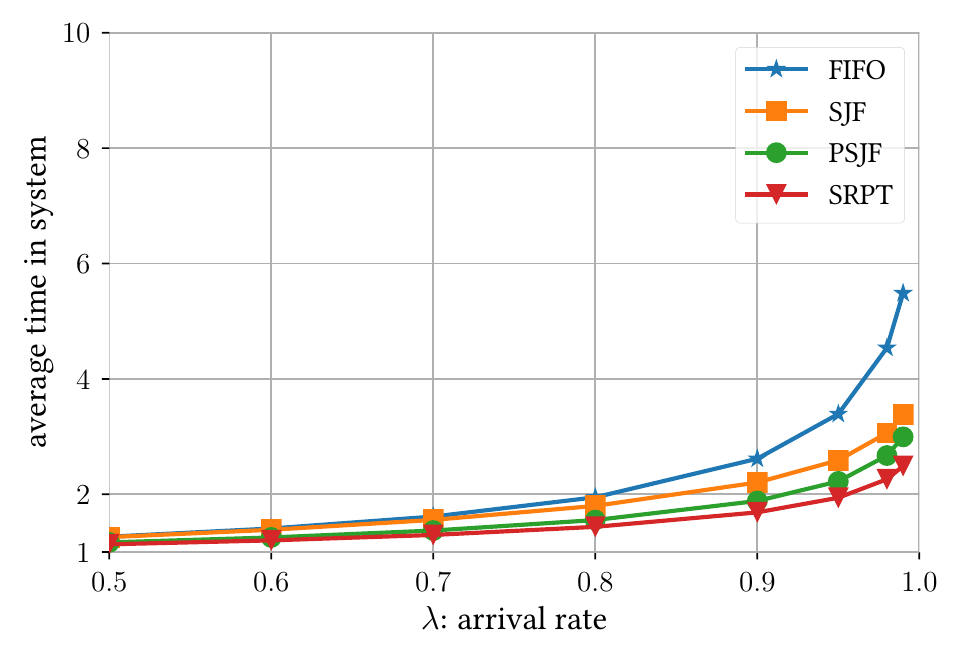}
    {Weibull service times, queue chosen by least loaded}{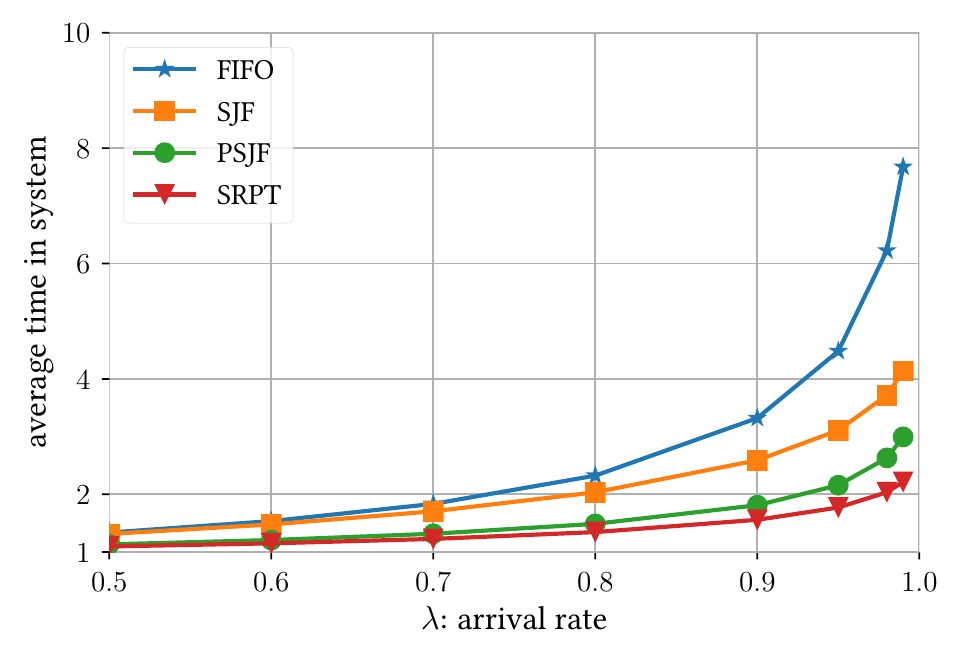}
    {Weibull service times, queue chosen by shortest queue}{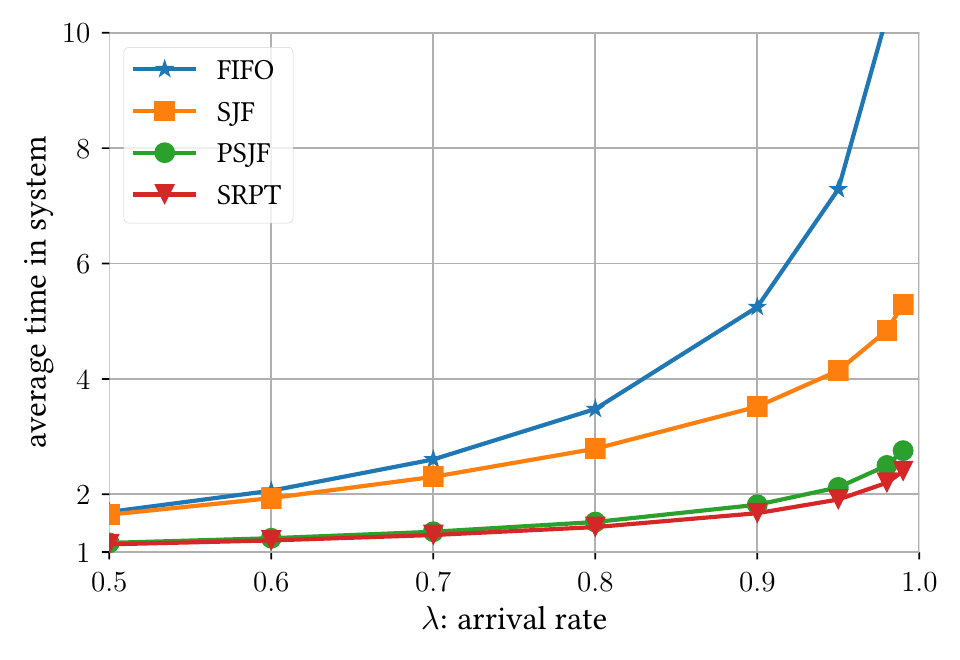}
    \caption{Exponential and Weibull service times, two choice supermarket model, various queue scheduling policies.}
    \label{fig:noprediction}
\end{figure*}

For single queues, Wierman and Nuyens look at variations of SRPT and SJF with inexact
\replaced{service time}{job size}s, bounding the performance gap based on bounds on how inexact
the estimates can be \cite{WiermanNuyens}.  Dell'Amico, Carra, and Michiardi
note that such bounds may be impractical, as outliers in estimating \replaced{service time}{job
size}s occur frequently;  they empirically study scheduling policies for
queueing systems with estimated \replaced{service time}{size}s \cite{DCM}.  We note \cite{DCM}
points out there are natural methods to estimate \replaced{service time}{job size}, such as by
running a small portion of the code in a coding job;  we expect this
or other inputs would be features in a\added{n} ML formulation.
Recent work by Scully and Harchol-Balter have considered 
scheduling policies that are based on the amount of service received, where the
scheduler only knows the service received approximately, subject to adversarial noise,
and the goal is to develop robust policies \cite{scully2018}.  
Also, for single queues, many prediction-based policies appear to fit within the more general framework of SOAP policies presented by Scully et~al.~\cite{scully2018also}.  

\added{
Works discussed above tackle the problem of evaluating when policies based on predicted \replaced{service time}{size} are beneficial, in the sense that they outperform non size-based approaches. In the context of this study, we refer especially to works that evaluate analytically~\cite{mitz2019scheduling} and experimentally~\cite{dellamico2019scheduling} the scheduling policies we use; in particular, they find that PSPJF (referred to as SPTE in~\cite{dellamico2019scheduling}) is quite robust against \replaced{service time}{job size} estimation errors, being outperformed by non size-based approaches only in extreme cases of large estimation errors and high service time skew.
}

Our work differs from these past works, in providing a model
specifically geared toward studying performance with machine-learning
based predictions in the context of the supermarket model.

\section{Known Service Times} 

\subsection{Scheduling Beyond FIFO}

To begin, we note again that the work
of \cite{DBLP:journals/pomacs/HellemansH18} shows that for the
supermarket model with $d$ choices for constant $d$, known service
times (independently chosen from a given service time distribution),
and FIFO scheduling, the equations for the stationary distribution can
be determined, when the queue is chosen according to the least loaded
policy. However, there appears to previously not have been studies of
scheduling schemes within each queue that make use of the service
times, including shortest job first (SJF), 
preemptive shortest job first
(PSJF), and  
shortest
remaining processing time (SRPT).  

While primarily in this paper we are inter\added{e}sted in the performance
of the supermarket model with {\em predicted} service times,
as these variations do not appear to have been studied, we provide results 
as a baseline for our later results.

\begin{table*}[thbp]
\begin{center}
\caption{Results from choosing from the shortest queue
compared with choosing the least loaded.}
\label{tab:table1}
\begin{tabular}{l rrrr rrrr}
\toprule
\multirow{2}{*}{$\lambda$} & \multicolumn{4}{c}{Shortest queue} & \multicolumn{4}{c}{Least loaded} \\
 & FIFO & SJF & PSJF & SRPT & FIFO & SJF & PSJF & SRPT  \\
\cmidrule(r){1-1} \cmidrule(lr){2-5} \cmidrule(l){6-9}
0.5 &  1.2658 & 1.2585 & 1.1669 & 1.1337 & 1.1510 & 1.1460 & 1.1462 & 1.0973 \\
0.6 &  1.4078 & 1.3857 & 1.2527 & 1.2020 & 1.2401 & 1.2280 & 1.2289 &  1.1518 \\
0.7 &  1.6148 & 1.5567 & 1.3726 & 1.2962 & 1.3749 & 1.3467 & 1.3490 & 1.2307 \\
0.8 &  1.9485 & 1.7997 & 1.5542 & 1.4367 & 1.5975 & 1.5297 & 1.5371 & 1.3533 \\
0.9 &  2.6168 & 2.2054 & 1.8850 & 1.6873 & 2.0534 & 1.8634 & 1.8915 & 1.5783 \\
0.95 &  3.3923 & 2.5903 & 2.2248 & 1.9408 & 2.5852 & 2.1999 & 2.2685 & 1.8096 \\
0.98 &  4.5384 & 3.0618 & 2.6721 & 2.2614 & 3.3798 & 2.6197 & 2.7807 & 2.1038 \\
0.99 &  5.4855 & 3.3856 &  2.8596 & 2.4903 & 4.0451 & 2.8514 & 3.1696 & 2.3176 \\
\bottomrule
\end{tabular}
\end{center}
\end{table*}

\begin{figure*}[thbp]
  \centering
  \twofig
    {Exponential service times, queue choice methods}{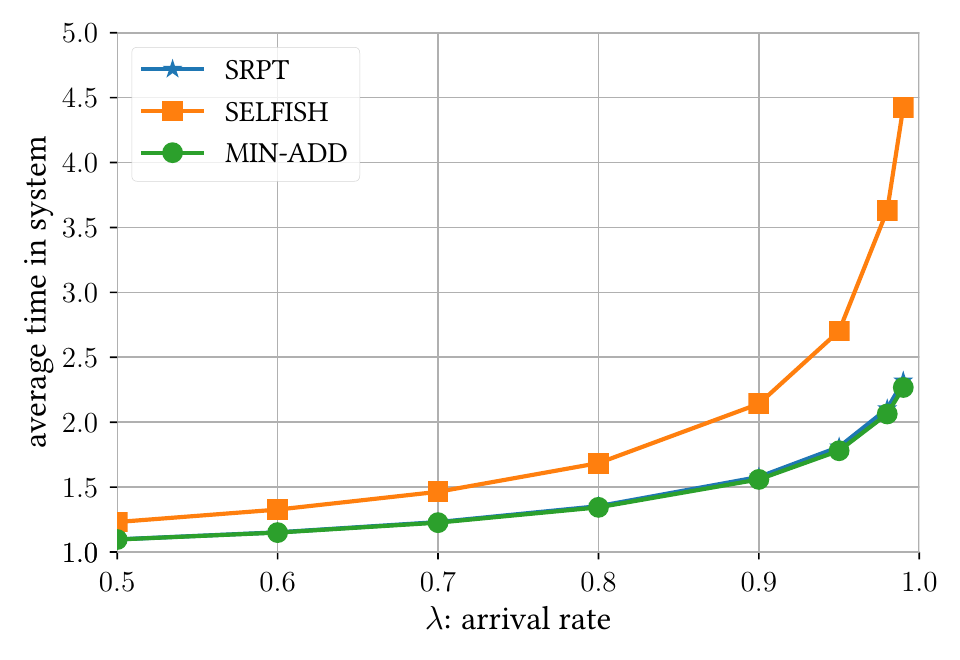}
    {Weibull service times, queue choice methods}{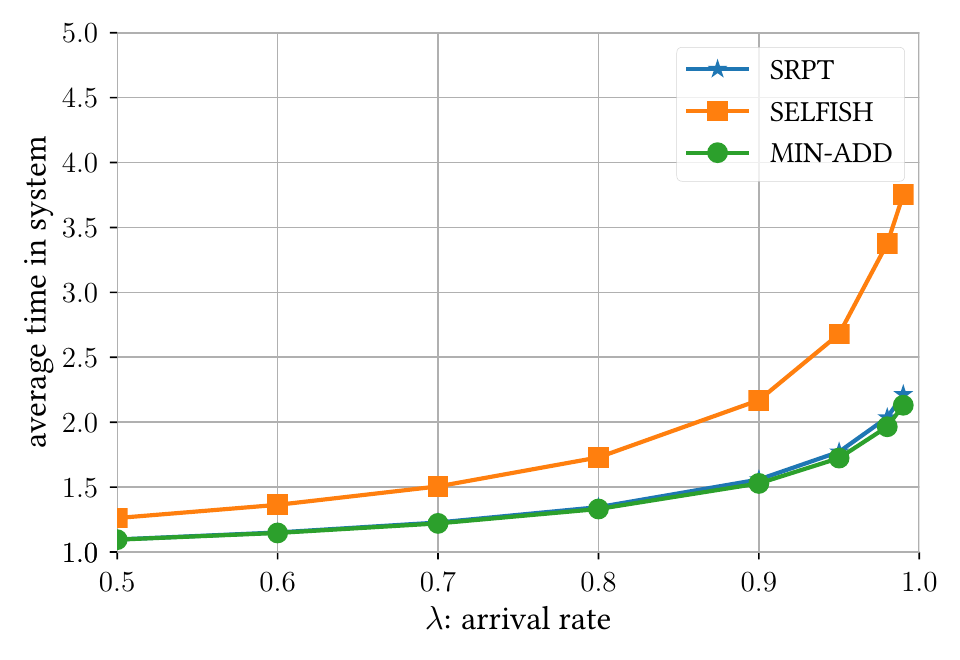}
\caption{Comparing methods of choosing a queue.  All queues use SRPT within the queue;  in the figure, SRPT means each job chooses the queue with smallest remaining work, SELFISH means each job chooses
the queue that minimizes its waiting time, and MIN-ADD means each job chooses the queue that minimizes the 
additional waiting time added.}
   \label{fig:selfmin}
\end{figure*}

In the simulation experiments we present, we simulate 1000
initially empty queues over 10000 units of time, and take the average
\replaced{time in system}{response time} for all jobs that terminate after time 1000 and before
time 10000.  We then take the average of this value over 100
simulations.  Waiting for the first 1000 time units allows the system
to approach the stationary distribution.  Variations of the
supermarket model have a limiting equilibrium distribution as the
number of queues goes to infinity
\cite{bramson2010randomized}, and in practice we find 1000 queues provides
an accurate estimate of the limiting behavior.  In the experiments we
focus on two example service distributions: exponential with mean 1,
and a Weibull distribution with cumulative distribution
$1-e^{-\sqrt{2x}}$.  (The Weibull distribution is more heavy-tailed,
but also has mean 1.  We have also done experiments with a more
heavy-tailed Weibull distribution with cumulative distribution
$1-e^{-\sqrt[3]{6x}}$; the general trends are similar for this
distribution as for the Weibull distribution we discuss.)  Arrivals
are Poisson with arrival rate $\lambda$; we focus on results with
$\lambda \geq 0.5$, as for smaller arrival rates all our proposed
schemes perform very well and it becomes difficult to see performance
differences.  Unless otherwise noted in the simulations each job
chooses $d=2$ queues at random.  While we have done simulations for
larger $d$ values, and at a high level there are similar trends,
studying the detailed effects of larger $d$ across the many variations
we study is left for future work.


Figure~\ref{fig:noprediction}(a) shows the results where the least
loaded queue is chosen (ties broken randomly), while
Figure~\ref{fig:noprediction}(b) shows the results where the shortest
queue is chosen, for exponential service times.
Figures~\ref{fig:noprediction}(c) and~\ref{fig:noprediction}(d)
present the results for the Weibull distributed service times.
Generally, we see that the results from using the known service times
to order jobs at the queue is very powerful; indeed, the gain from
using SRPT appears larger than the gain from moving from shortest
queue to least loaded, and similarly the gain from using SJF and PSJF
is larger under high enough loads.

As the charts make it somewhat more difficult to see some important details, we
present numerical results for exponentially distributed service times in
Table~\ref{tab:table1} to mark some key points.  While generally the
benefits from using the service times to both choose the queue and
order the queue are complementary, this is not always the case.  We
see that using least loaded rather than shortest queue when using PSJF
can increase the average time in the system under suitably high load.
(This also occurs with the Weibull distribution under sufficiently
high loads.)  We also see that using PSJF can give worse performance
than using SJF; however, this does not happen with our experiments
with the Weibull distribution, where the ability of preemption to help
avoid waiting for long-running jobs appears to be more helpful.  While
it is known that PSJF can behave worse than SJF, these examples
highlight that the interactions when using service time information in
multiple choice systems must be treated carefully.


%

%

\subsection{Choosing a Queue with Exact Information}

Given the improvements possible using known service times in the
supermarket model, we now consider methods for choosing a queue beyond the
queue with the least load, in the setting withou\deleted{o}t predictions.  Given full information about the service
times of jobs at each queue, a job could be placed so that it {\em
minimizes the additional waiting time}.  The additional waiting time
when placing an arriving job is the sum of the remaining service times
of all jobs in the queue that will remain ahead of the arriving job,
summed with the product of the service time of the arriving job and
the number of jobs it will be placed ahead of.  Equivalently, we can
consider the total waiting time for each queue before and after the
arriving job would be placed (ignoring the possibility of future
jobs), and place the item in the queue that leads to the smallest
increase.

Alternatively, if control is not centralized, we might consider {\em
selfish} jobs, that seek only to minimize {\em their own} waiting time
when choosing a queue.  In this case the arriving job will consider
the sum of the remaining service times of all jobs that will be ahead
of it for each available queue choice.

Our results, given in Figure~\ref{fig:selfmin}, show that choosing a queue to
minimize the additional waiting time in these situations does yield a
small improvement over least loaded SRPT, as might be expected.
Because the additional improvement is small, we expect in many systems
it may not be worthwhile to implement this modification, even if
expected waiting time is the primary performance metric.  Our results
also show that while selfish jobs have a significant negative effect,
the overall average service time still remains smaller than the
standard supermarket model when choosing the shorter of two FIFO
queues.

\section{Predicted Service Times} 

In many settings, it may be unreasonable to expect to obtain exact
service times, but predicted service times may be available.  Indeed,
with advances in ML techniques, we expect that in many
settings some type of prediction will be available.  As noted
in \cite{mitz2019scheduling}, in the context of scheduling within a
single queue, one would expect that even weak predictors may be very useful,
since ordering jobs correctly most of the time will produce
significant gains.  As we have seen, however, even without
predictions the question of whether using load information for both
choosing a queue and for ordering within a queue provides
complementary gains is not always clear.  Naturally, the same question
will arise again when using predicted service times. 

\subsection{The Prediction Model}

In what follows, we utilize a simple model used
in \cite{mitz2019scheduling}, namely that there is a continuous joint
distribution $g(x,y)$ for the actual service time $x$ and predicted
service time $y$.


With this model, there remains an issue of how to describe the predicted
remaining service time.  Suppose that the original predicted service
time for a job is $y$, but the actual service time is $x > y$.  If the
amount of service is being tracked, and the service received has been
$t$, then as the remaining service time is $x-t$, it is natural to use
$y-t$ as the predicted remaining service time.  Of course, at some
point it becomes the case that $t > y$, and the predicted remaining
service time will be negative, which seems unsuitable.

We use $(y-t)^+ = \max(y-t,0)$ as the predicted remaining service
time.  We recognize (as noted in \cite{mitz2019scheduling})
that this is problematic; clearly the predicted remaining service
time should be positive, and ideally would be a function $f(y,t)$ of
the initial prediction and the time served thus far.  However,
determining the appropriate function would appear to require some
knowledge of the joint distribution $g(x,y)$; our aim here is to
explore simple, general approaches (such as choosing the shortest of
two queues and using SRPT) that are agnostic to the underlying
distribution $g$.  In many situations, it may be computationally
undesirable to utilize knowledge of $g$, or $g$ may be not known or
changing over time.  We therefore leave the question of how to
optimize the estimate of the predicted remaining time to achieve the
best performance in this context as future work.

We consider various models for predictions (some of which were used
in \cite{mitz2019scheduling}).  The models are intended to be exemplary;
they do not match a specific real-world setting, and indeed it would 
be difficult to consider the range of possible real-world predictions.
Rather, they are meant to show generally that even moderately accurate
predictions can yield strong performance, and to show that a variety 
of interesting behaviors can occur under this framework.  

\begin{figure*}[t!]
  \centering
  \fourfig
    {Exponential service times, queue chosen by least loaded updated}{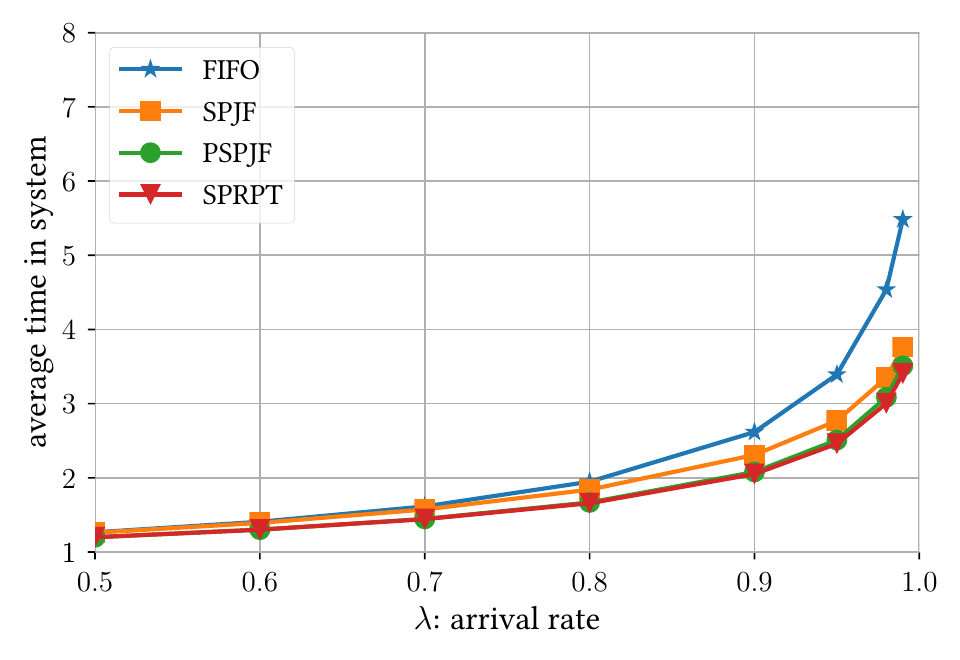}
    {Exponential service times, queue chosen by shortest queue}{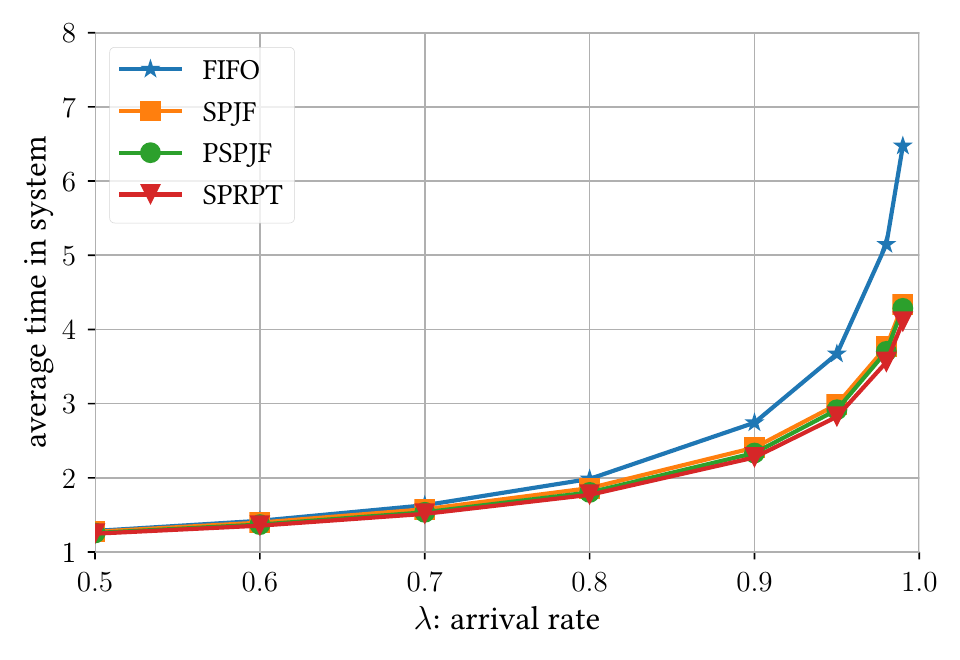}
    {Weibull service times, queue chosen by least loaded updated}{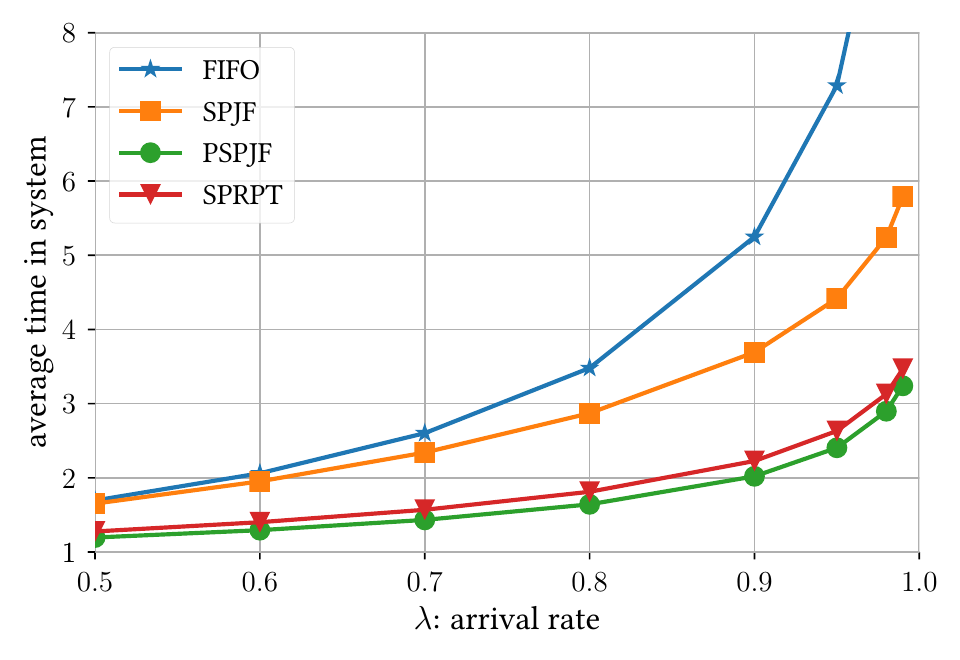}
    {Weibull service times, queue chosen by shortest queue}{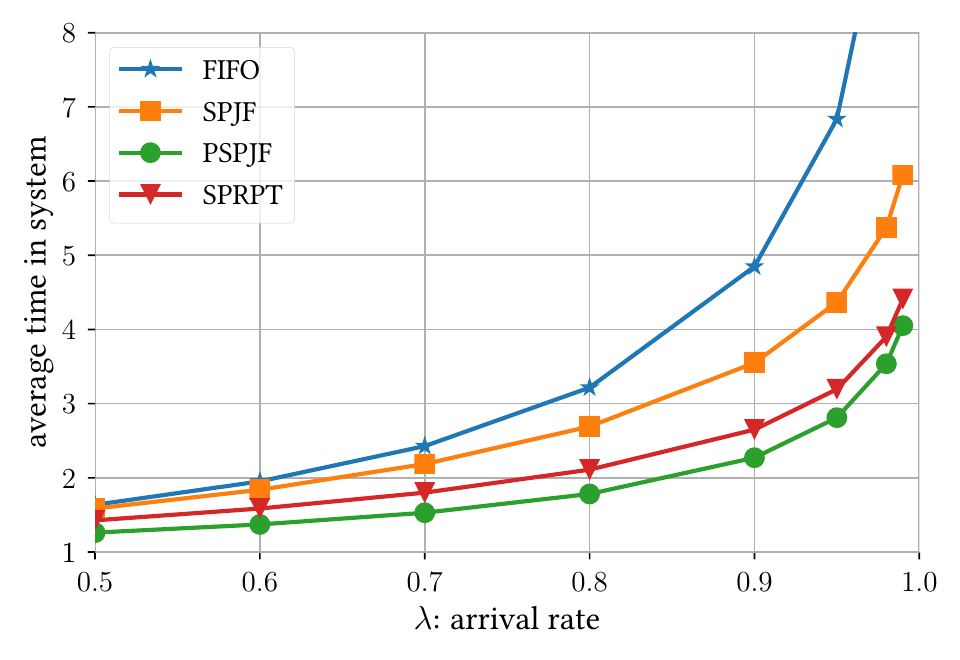}
  \caption{Exponential predictions with exponential and Weibull service times, two choice supermarket model, with various queue scheduling policies.}
  \label{fig:expprediction}
\end{figure*}

In one model, which we refer to as {\em exponential predictions}, a
job with actual service time $x$ has a predicted service time that is
exponentially distributed with mean $x$.  This model is not meant to
accurately represent a specific situation, but is potentially useful
for theoretical analysis in that the corresponding density equation
$g(x,y)$ is easy to write down, \added{has a non-zero probability for even very large prediction errors (note that the predicted service time will be any given constant factor larger than the true service time with constant probability)}, and it highlights how even noisy \added{(and sometimes extremely incorrect)}
predictions can perform well.  Also, exponential service times are a
standard first consideration in queueing theory.  In another model,
which we refer to as {\em $\alpha$-predictions}, a job with service
time $x$ has a predicted service time that is uniform over
$[(1-\alpha)x,(1+\alpha)x]$, for a scale parameter $0 \leq \alpha \leq
1$.  Again, this is a simple model that captures inaccurate estimates
naturally.  Finally, we introduce a model that we dub {\em
$(\alpha,\beta)$-predictions}, which makes use of the following notion
of a {\em reversal}.  For a service distribution with \deleted{be} the
cumulative distribution function $S(x)$, the reversal of $x$ is
$S^{-1}(1-S(x))$.  For example, if $x$ is the value that is at the
70th percentile of the distribution, the reversal is the value at the
30th percentile of the distribution.  For an {\em
$(\alpha,\beta)$-prediction}, when the service time is $x$, with
probability $\beta$ we return the reversal of $x$, and with all
remaining probability the predicted service time is uniform over
$[(1-\alpha)x,(1+\alpha)x]$.  We use this model to represent cases
where severe mispredictions are possible, so that jobs with very large
service times might be mistakenly predicted as having very small
service times (and vice versa).  We might expect such mispredictions
could be potentially very problematic when scheduling jobs according
to their predicted service times.

\begin{figure*}[t!]
  \centering
  \fourfig
    {Exponential service times, queue chosen by least loaded updated}{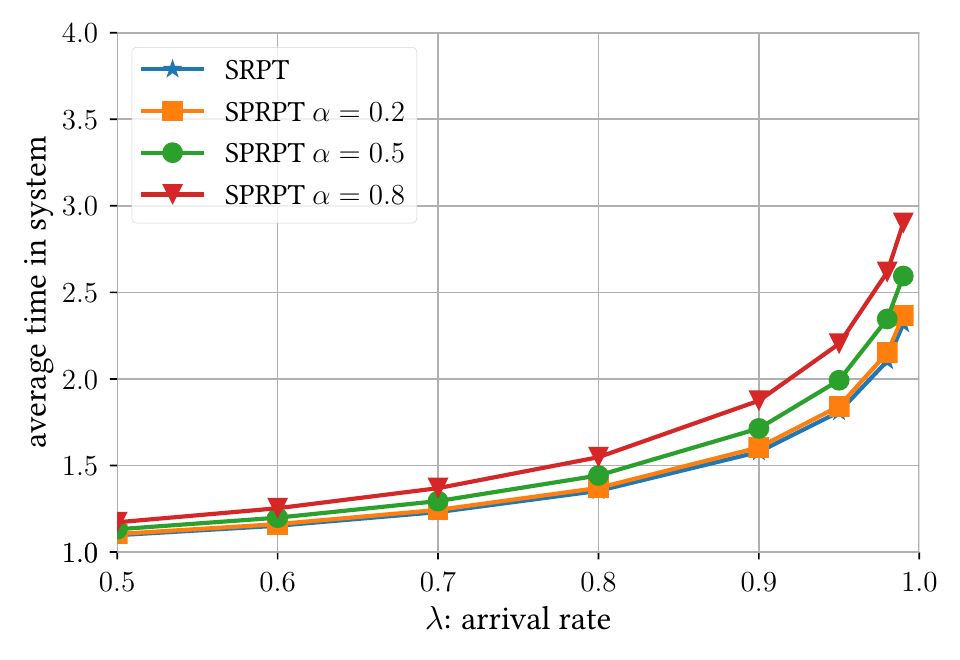}
    {Exponential service times, queue chosen by shortest queue}{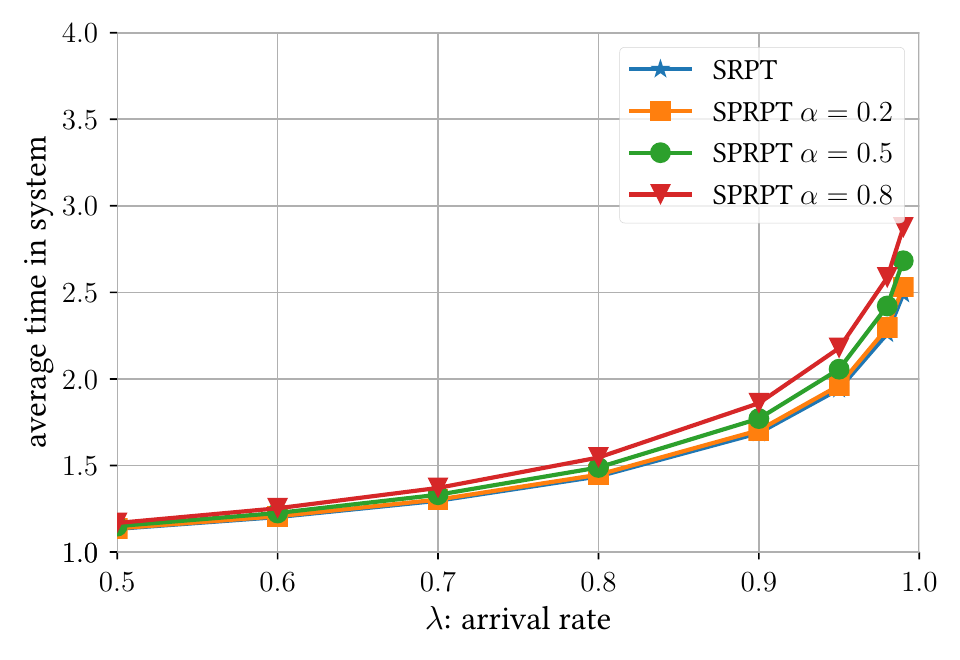}
    {Weibull service times, queue chosen by least loaded updated}{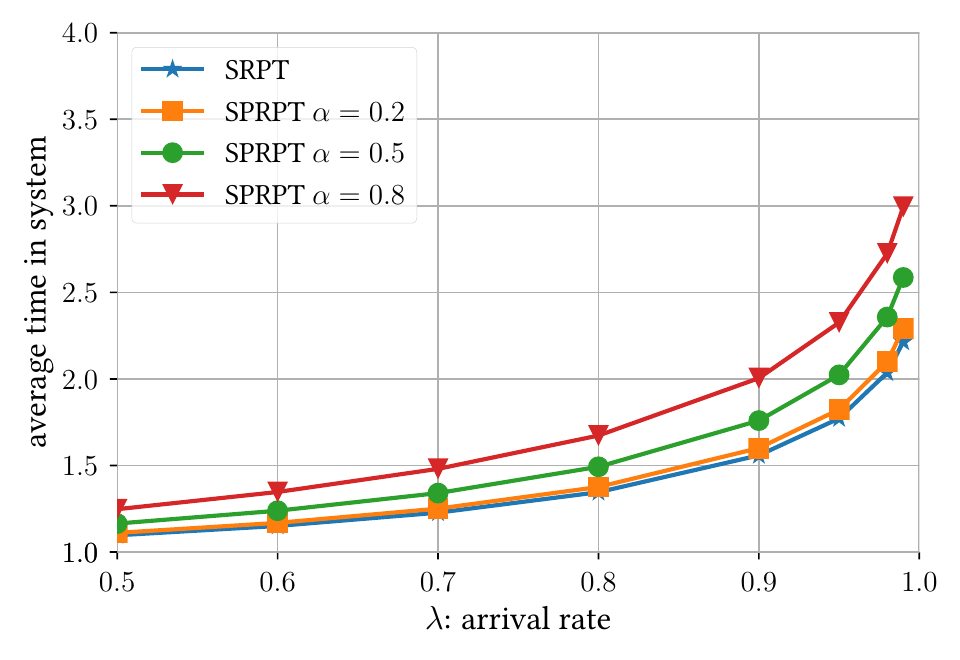}
    {Weibull service times, queue chosen by shortest queue}{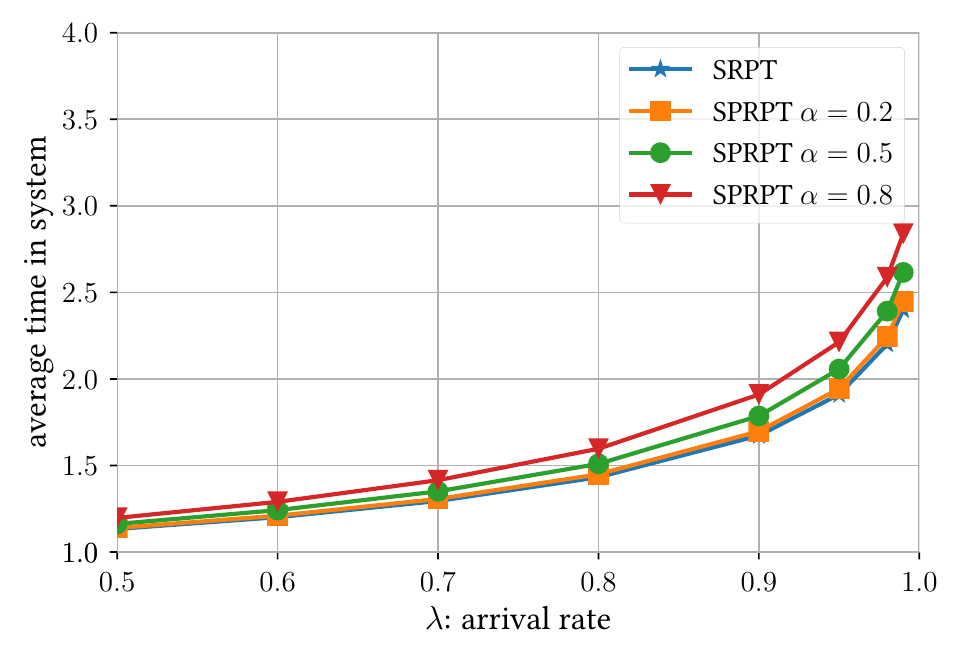}
\caption{$\alpha$-predictions with exponential and Weibull service times, two choice supermarket model, using SPRPT.}
    \label{fig:alphaprediction1}
\end{figure*}

There \deleted{further} remains the question of how to account for the
predicted workload at a queue.  We discuss several variations.
\begin{enumerate} 
\item {\bf Least Loaded Total:}  One could
simply treat the predicted service times as actual service times,
and track the total predicted service time remaining at a queue.
That is, when a new job arrives at a queue, the predicted service time
for the job is added to the total, and the total predicted service
time reduces at a rate of one unit per unit time when a job is in the
system (with a lower bound of 0);  when the queue empties, the total predicted service time
is reset to 0.  An advantage of this approach is that in implementation 
the queue state can be represented by a single number.  The disadvantage
is that when a job's predicted service time differs greatly from the
real service time, this approach does not correspondingly update when
that job completes. 
\item {\bf Least Loaded Updated:}  Here one updates the queue state both on a job arrival
and a job completion;  when a job completes, the predicted service time
at the queue is recomputed as the sum of the predicted service times of the
remaining jobs.  With small additional complexity, the accuracy of the predicted work
at the queue improves substantially.  
\item {\bf Shortest Queue:} One can always simply use the number of jobs rather than the 
predicted service time to choose the queue.  
\end{enumerate} 
We note Least Loaded Updated performs much better than Least Loaded Total, and 
focus on it for the rest of the paper.  More on the comparison is given in 
\cref{app:cq2}.

\begin{figure*}[t!]
  \centering
  \fourfig
    {Exponential service times, queue chosen by least loaded updated}{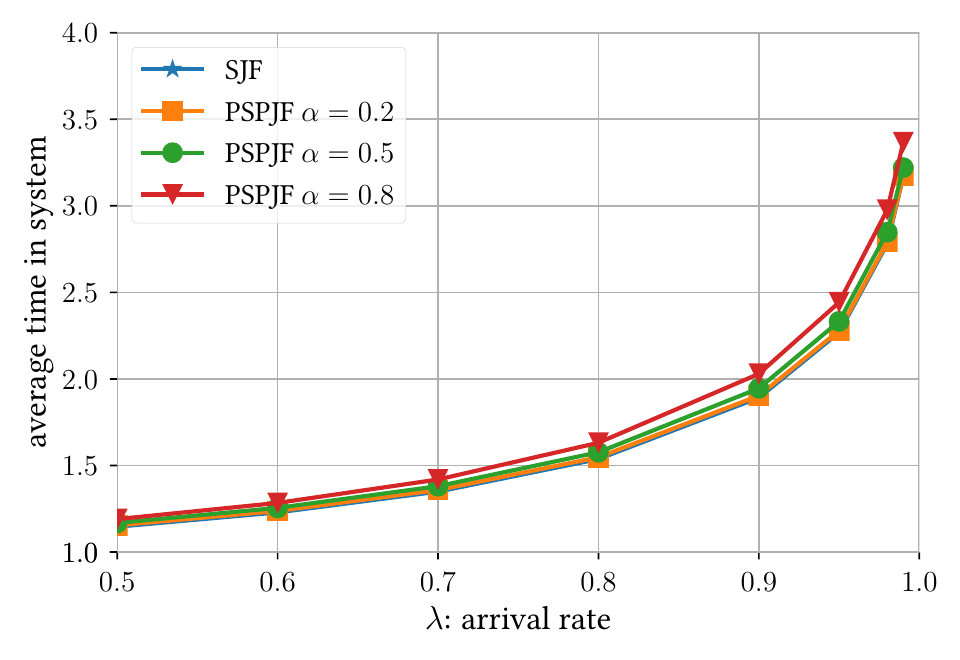}
    {Exponential service times, queue chosen by shortest queue}{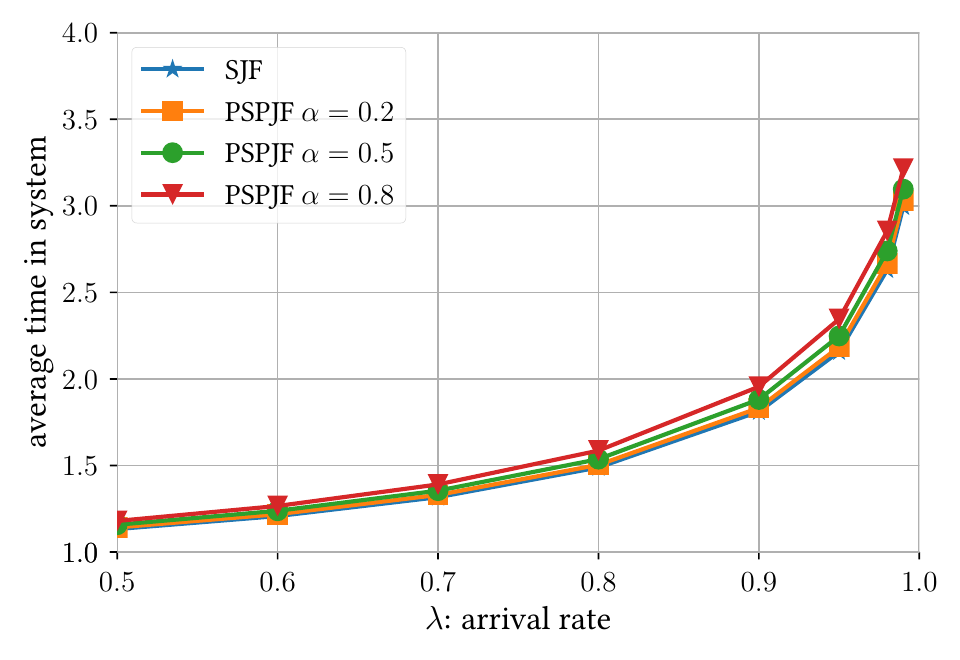}
    {Weibull service times, queue chosen by least loaded updated}{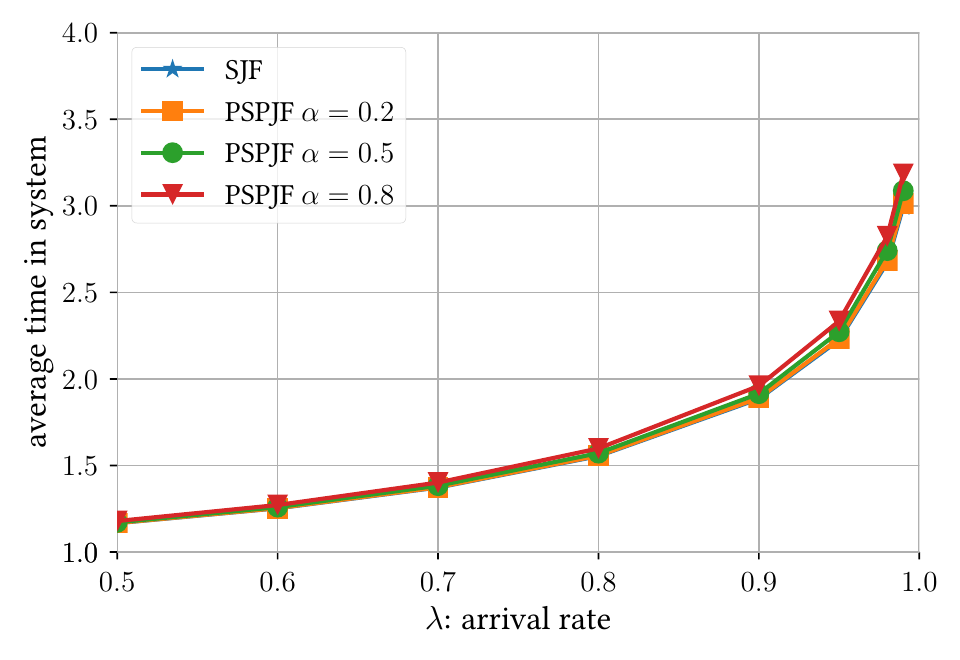}
    {Weibull service times, queue chosen by shortest queue}{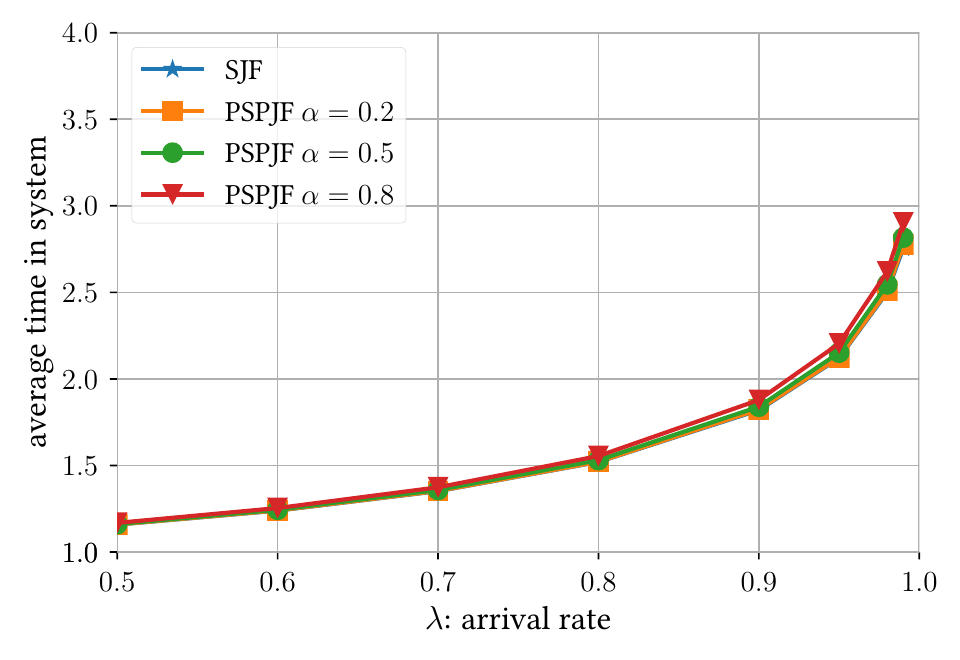}
\caption{$\alpha$-predictions with exponential and Weibull service times, two choice supermarket model, using PSPJF.}
    \label{fig:alphaprediction2}
\end{figure*}

\begin{figure*}[t!]
  \centering
  \fourfig
    {Exponential service times, queue chosen by least loaded updated}{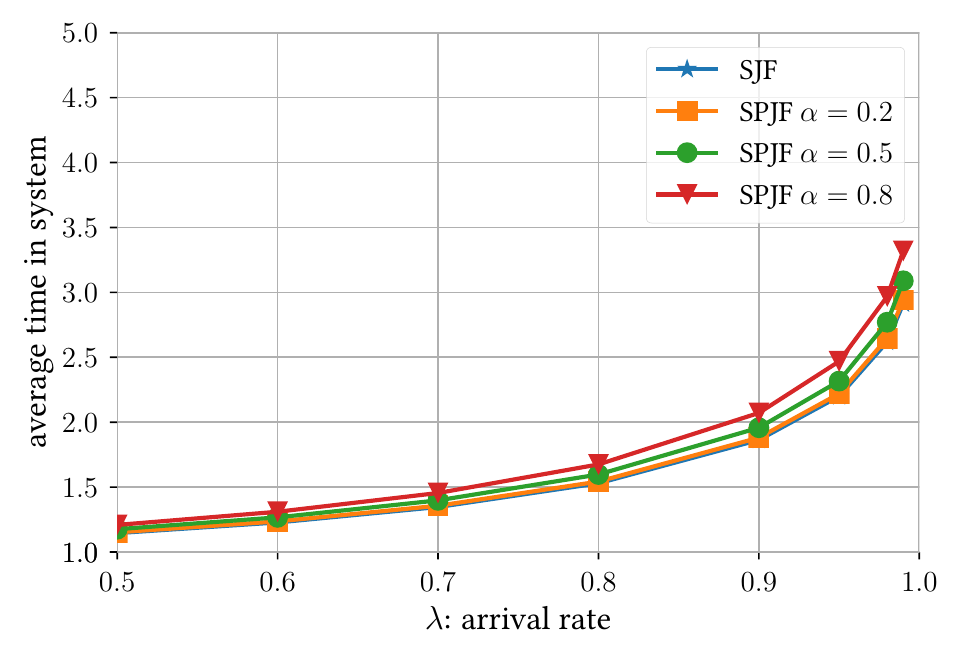}
    {Exponential service times, queue chosen by shortest queue}{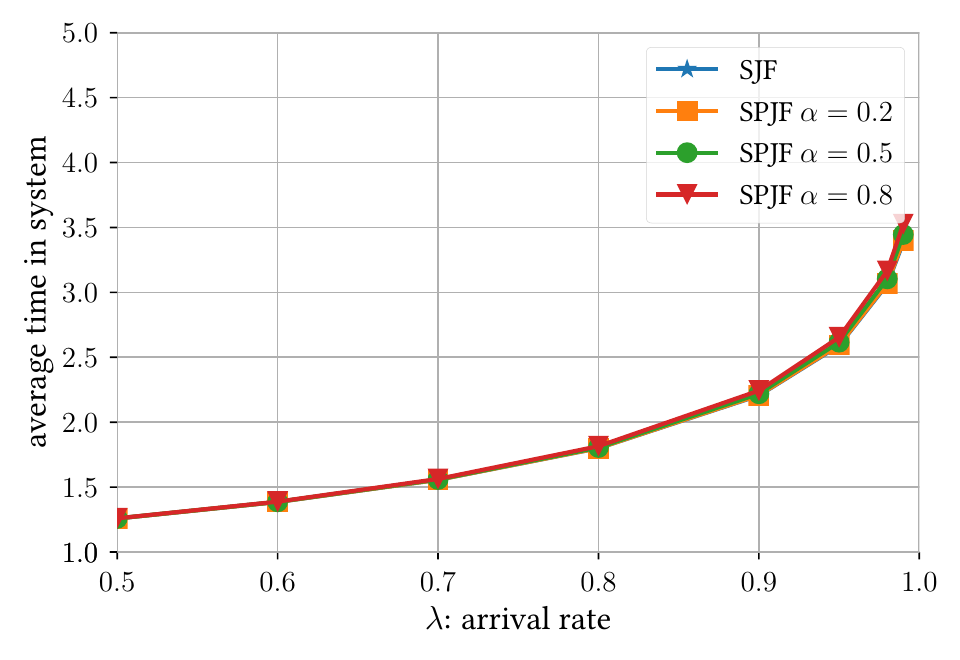}
    {Weibull service times, queue chosen by least loaded updated}{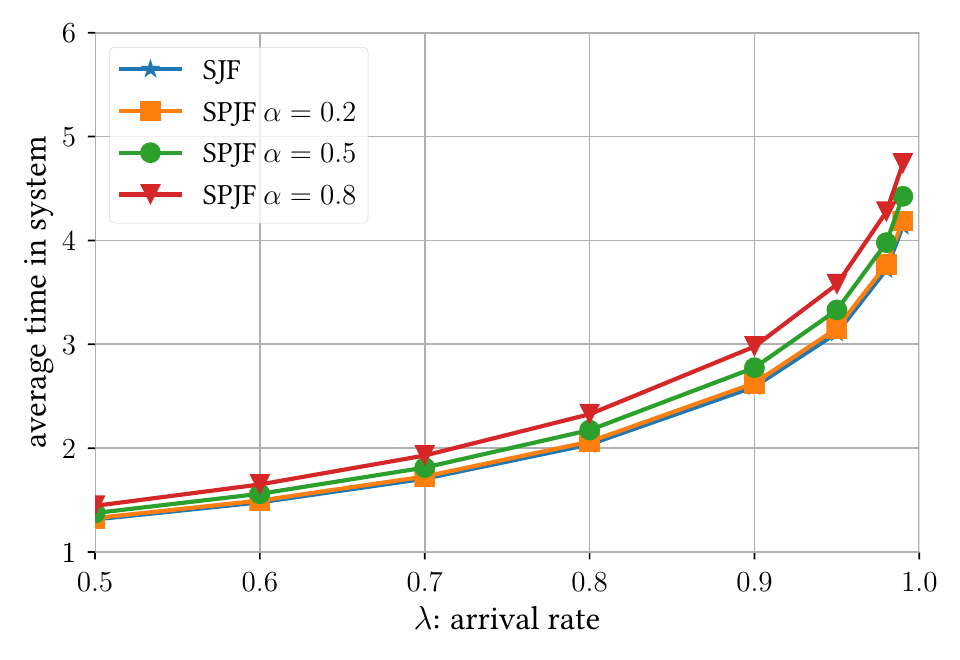}
    {Weibull service times, queue chosen by shortest queue}{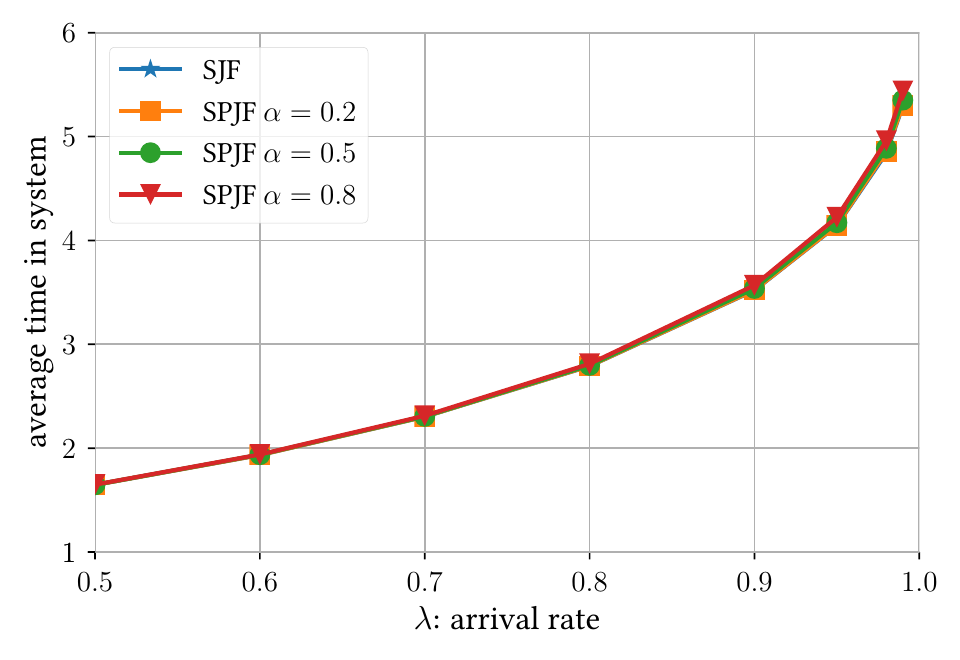}
\caption{$\alpha$-predictions with exponential and Weibull service times, two choice supermarket model, using SPJF.}
    \label{fig:alphaprediction3}
\end{figure*}

\begin{figure*}[t!]
  \centering
  \fourfig
    {Exponential service times, queue chosen by least loaded updated}{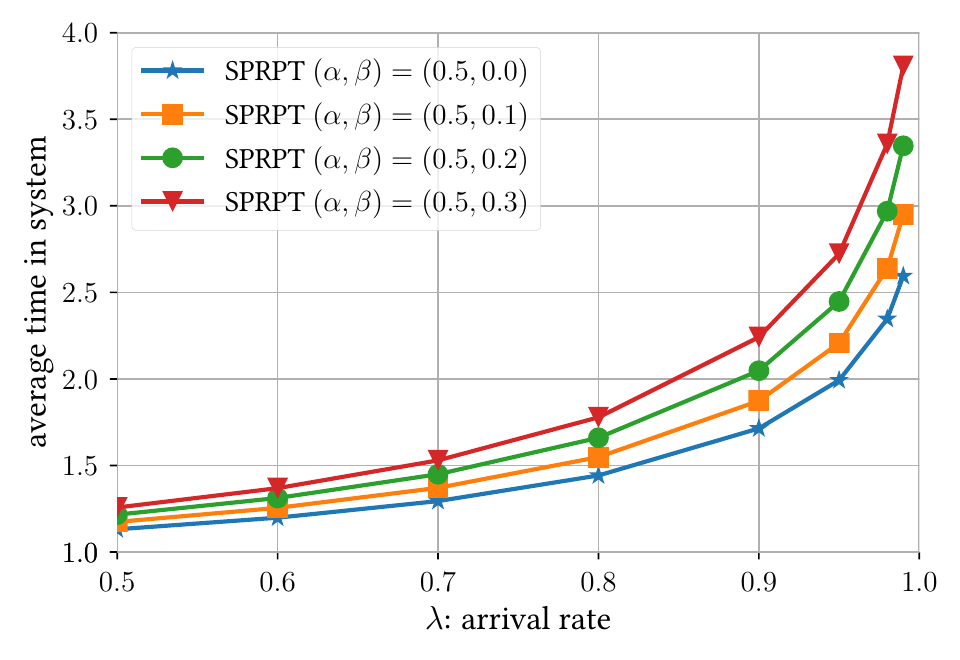}
    {Exponential service times, queue chosen by shortest queue}{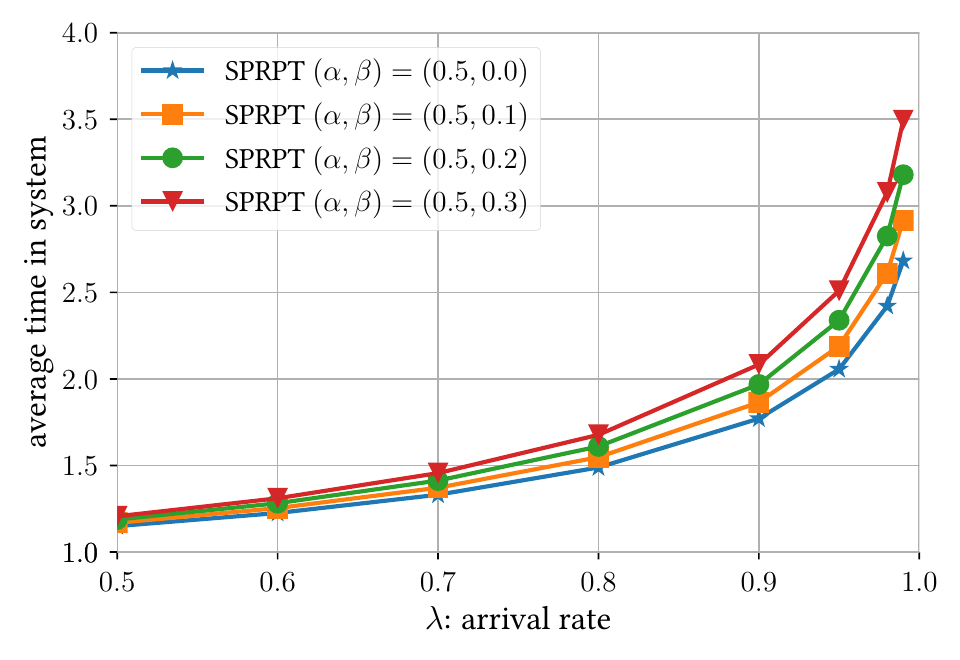}
    {Weibull service times, queue chosen by least loaded updated}{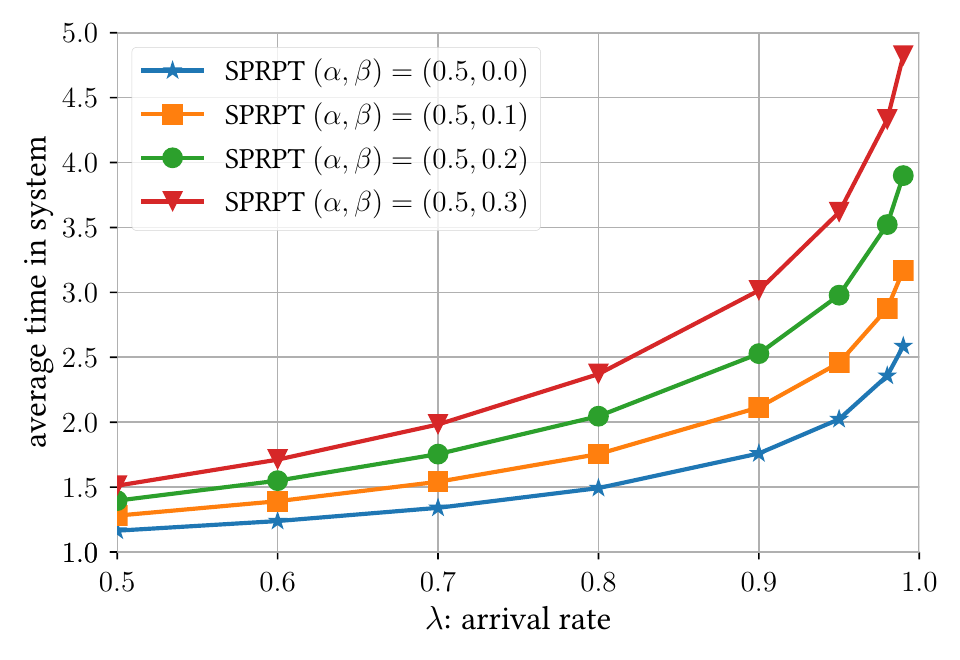}
    {Weibull service times, queue chosen by shortest queue}{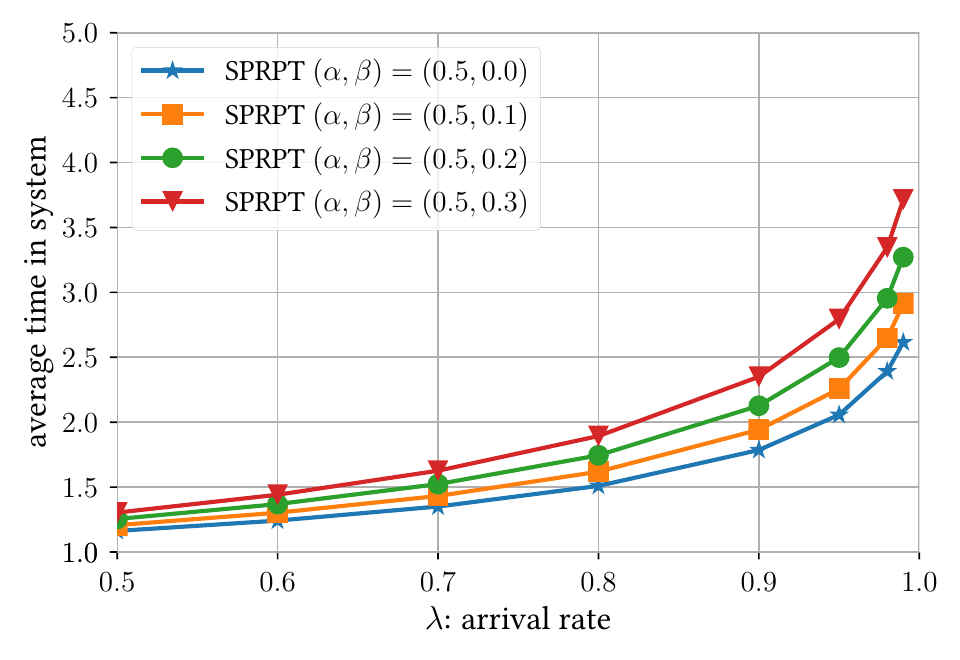}
\caption{$(\alpha,\beta)$-predictions with exponential and Weibull service times, two choice supermarket model, using SPRPT.}
    \label{fig:alphabetaprediction1}
\end{figure*}

\begin{figure*}[t!]
  \centering
  \fourfig
    {Exponential service times, queue chosen by least loaded updated}{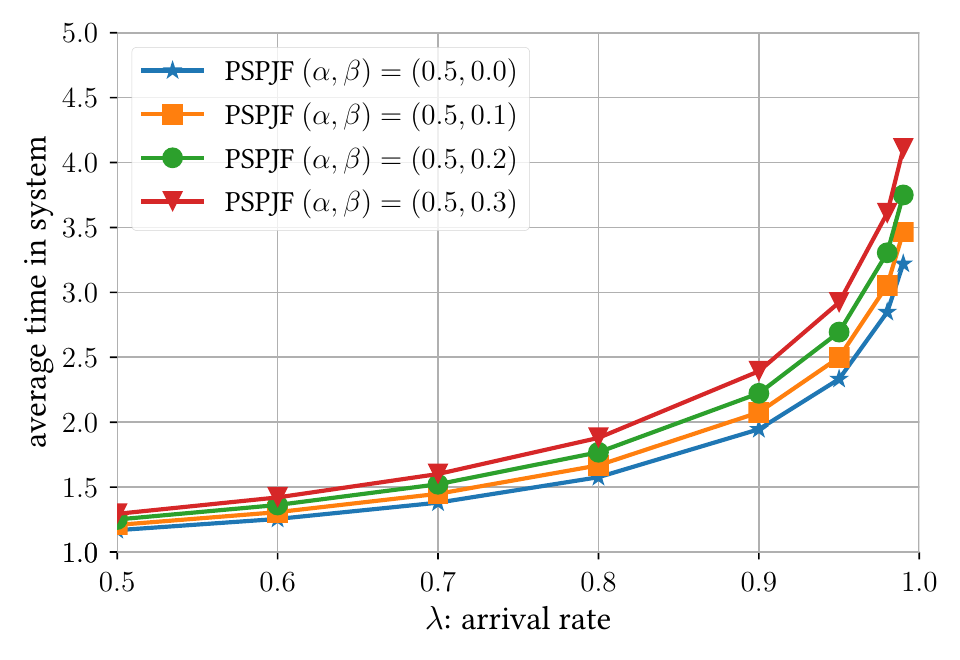}
    {Exponential service times, queue chosen by shortest queue}{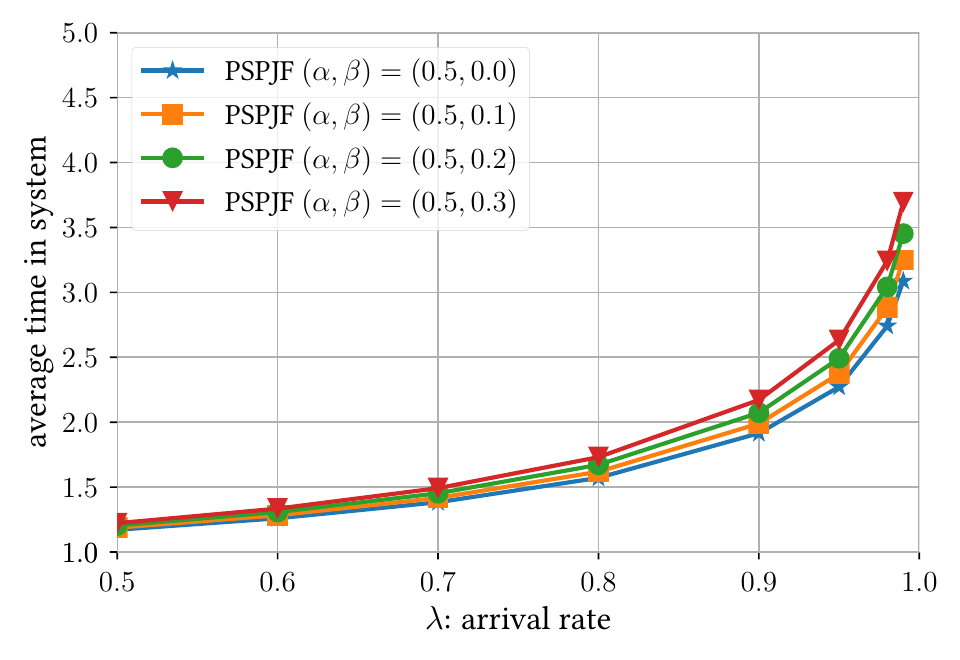}
    {Weibull service times, queue chosen by least loaded updated}{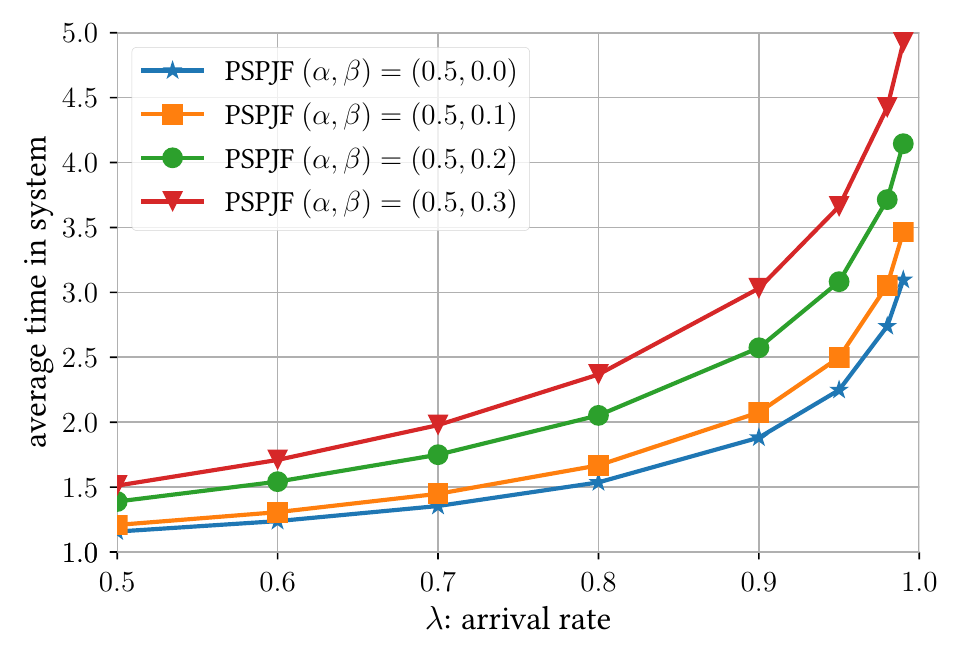}
    {Weibull service times, queue chosen by shortest queue}{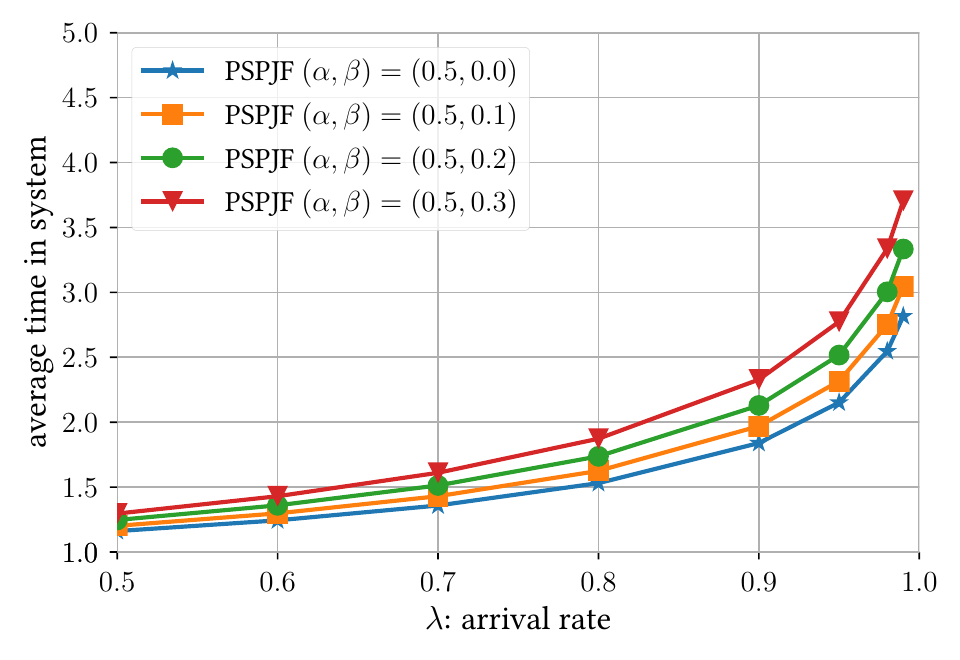}
\caption{$(\alpha,\beta)$-predictions with exponential and Weibull service times, two choice supermarket model, using PSPJF.}
    \label{fig:alphabetaprediction2}
\end{figure*}

\subsection{Scheduling with Predictions}

We begin as before by first considering the effect of the choice of
scheduling procedure within a queue, by examining results for FIFO,
shortest predicted job first (SPJF), preemptive shortest predicted job
first (PSPJF), and shortest predicted remaining processing time
(SPRPT) in various settings.  Our figures consider the 
least loaded updated and shortest queue
variations described above (as the least loaded total variation generally performs significantly worse, and so we do not generally consider this model,
although for comparison purposes we present some results for it in the next subsection).  We again consider exponential and Weibull
distributed service times as previously described.

Our first results for exponential predictions, shown in Figure~\ref{fig:expprediction},
already show two key points: predicted service
times can work quite well, but there are also surprising and
interesting behaviors.  First, choosing the shortest queue generally
performs \replaced{quite similarly to}{better than} choosing the least loaded according to the
predicted service times of jobs in the queue; for this set of
experiments, \replaced{choosing the queue based on its predicted load still slightly outperforms using just the number of jobs, but we will see this is not the case in other settings}{only with Weibull distributed service times and FIFO
service does choosing the queue based on the predicted load in the
queue perform better than using just the number of jobs}.  That is,
when using strategies within the queue that utilize the predicted
information, it \replaced{can be}{is} worse to use the predicted load to choose the
queue.  Hence\deleted{, even in this very simple case, we see that} using
predicted information for multiple subtasks (choosing a queue, and
balancing within a queue) \added{may not be beneficial and} can \added{even} lead to worse performance than simply
using the information for one of the subtasks.
Second, PSPJF performs better than SPRPT on the Weibull distribution.
On reflection, this seems reasonable from first principles; a long job that is
incorrectly predicted to have a small remaining processing time can
lead to increased waiting times for many jobs under SPRPT, but
preempting based on the initial prediction of the job time ameliorates
this effect.

We now examine results in the setting of $\alpha$-predictions.  
We first look at the case of SPRPT; results for other schemes have
similar characteristics.  We compare SRPT (no prediction) with SPRPT
for $\alpha = 0.2, 0.5,$, and $0.8$, both using the least loaded
update and shortest queue policies.  The results appear in 
Figure~\ref{fig:alphaprediction1}.\footnote{We note that in some of our plots throughout this section, the lines for SRPT or SJF are hard to distinguish, as they are very close to other plotted results.}
The primary takeaway is that
again using predictions offers what is arguably surprisingly little
loss in performance, even at large values of $\alpha$.  Here, we find
that least loaded does better than shortest queue for \replaced{exponential service times}{small values of
$\alpha$}, but for \added{the more skewed Weibull distribution and} $\alpha = 0.8$ \deleted{and high arrival rates} shortest queue
can perform slightly better.  \replaced{With large errors and skewed \replaced{service time}{job size} distributions}{This is consistent with our results for
the exponential model;  under large error}, using predictions both to choose
a queue and within a queue can lead to over-using the predictions \added{and worse performance}.

We also look at the case of PSPJF in Figure~\ref{fig:alphaprediction2}.
Performance is somewhat worse than for SPRPT, and the effect of
increasing $\alpha$ is smaller.  Here, we find that joining
the shortest queue generally does better than joining the least loaded
queue.  \deleted{\replaced{This}{Again, this} is consistent with our results for \replaced{SPRPT}{the exponential
model}.}  Overall the picture remains very similar.

For completeness we also provide results using SPJF\added{, which generally performs worse than SPRPT and PSPJF,}
in Figure~\ref{fig:alphaprediction3}.  \deleted{Here SPJF generally performs worse
than SPRPT and PSPJF; however} \added{However}, the effect on performance as $\alpha$ increases
rises even more slowly with $\alpha$.  In these experiments,
using least loaded always performed better than choosing the shortest queue.



\begin{figure*}[htbp]
  \centering
  \fourfig
    {Exponential service times, queue chosen by least loaded updated}{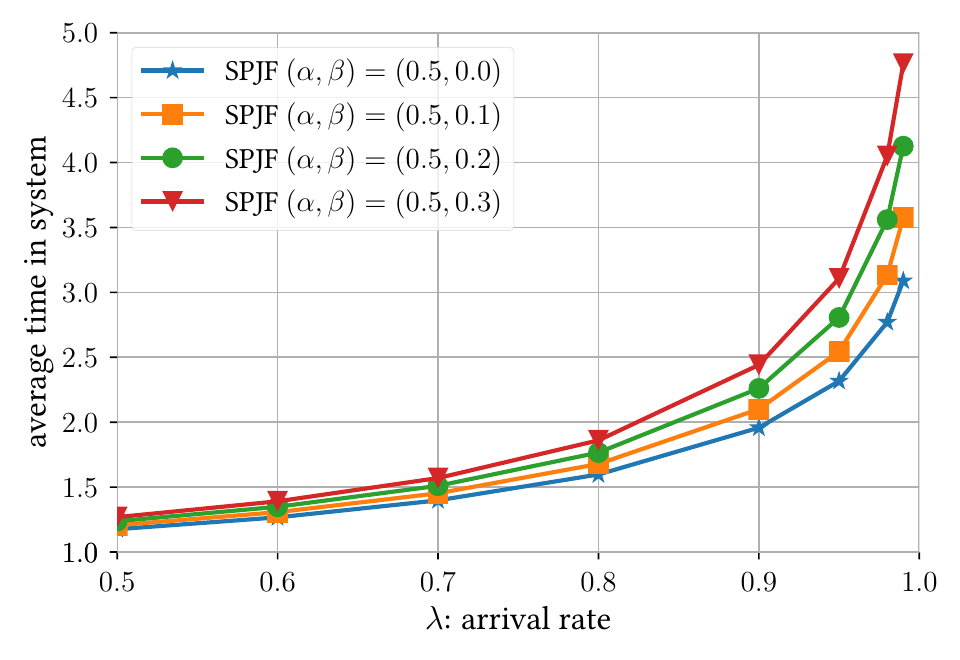}
    {Exponential service times, queue chosen by shortest queue}{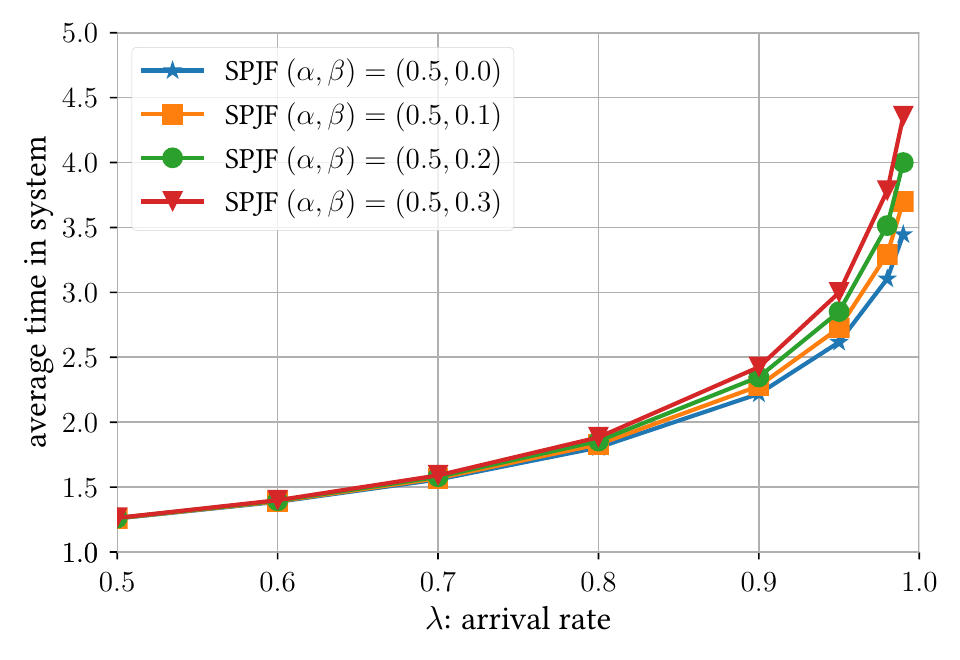}
    {Weibull service times, queue chosen by least loaded updated}{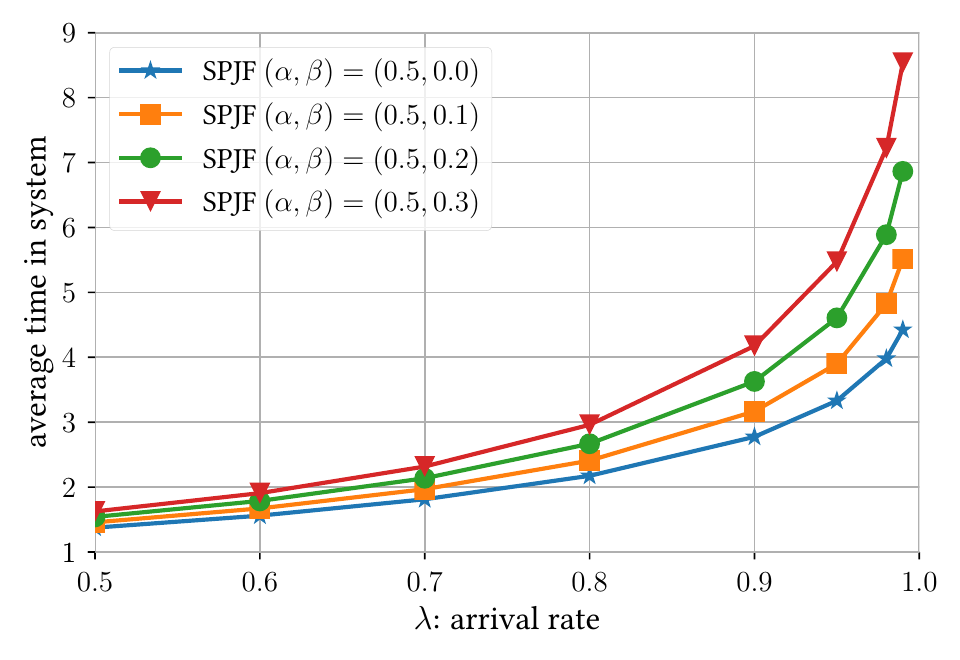}
    {Weibull service times, queue chosen by shortest queue}{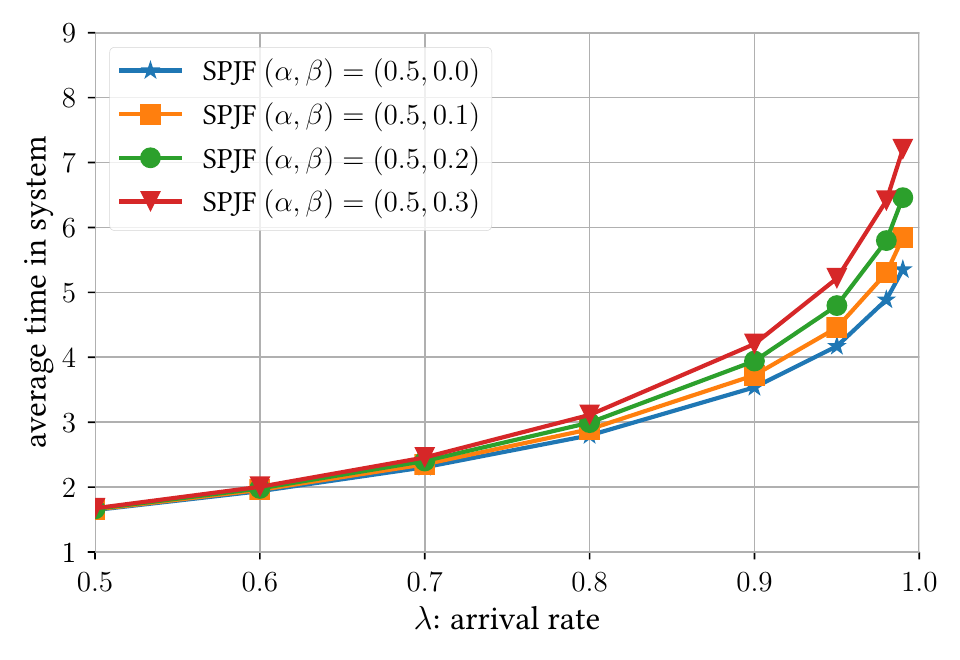}
\caption{$(\alpha,\beta)$-predictions with exponential and Weibull service times, two choice supermarket model, using SPJF.}
    \label{fig:alphabetaprediction3}
\end{figure*}

Finally, we consider the case of $(\alpha,\beta)$-predictions.  Here
we present an example of $\alpha = 0.5$ with $\beta = 0.1, 0.2$, and
$0.3$, comparing also with the results from $\alpha$-prediction when
$\alpha$ = 0.5.  Recall that in this setting with probability $\beta$
a job's service time is replaced by its reversal in the cumulative
distribution function, so that jobs with very large service times
might be mistakenly predicted as having very small service times (and
vice versa).  The remaining jobs have predictions uniform over
$[(1-\alpha)x,(1+\alpha)x]$ when the true service time is $x$.  
The results for SPRPT are given in Figure~\ref{fig:alphabetaprediction1},
for PSPJF are given in Figure~\ref{fig:alphabetaprediction2},
and for SPJF are given in Figure~\ref{fig:alphabetaprediction3}.
The
primary takeaway is again that performance is quite robust to mispredictions.
Even when $\beta = 0.3$, performance in all cases is significantly better than for standard choosing the shortest of two queues and using FIFO queueing without knowledge of service times.  We also see now familiar trends.  The effects of misprediction are more significant for the heavy-tailed service times, and when mispredictions are sufficiently frequent, it becomes better to choose a queue according to the shortest queue rather than according to the least loaded updated policy.  Also, in some cases PSPJF can outperform SPRPT\added{ (compare, e.g., \cref{fig:alphabetaprediction1}(c) and \cref{fig:alphabetaprediction2}(c) with $\beta=0.2$ and $\lambda \leq 0.8$)}.  

%

\subsubsection{Fairness Issues}
\label{sec:fairness}

As the general problem of utilizing predictions for queue scheduling
is relatively new, we have focused here on examining expected response
time.  We point out, however, that there are novel problems regarding
questions such as fairness in the setting where predictions are used.
For example, a standard notion of fairness involves considering job
\emph{slowdown}, i.e., the ratio $T(x)/x$ between a job $j$'s response
time $T(x)$ and its \replaced{service time}{size} $x$, and the \emph{mean conditional slowdown} $E[T(x)]/x$
\cite{DBLP:conf/sigmetrics/BansalH01,wierman2011fairness,WiermanMHB}.
Not surprisingly, we find that when using predictions, even when using
scheduling methods based on SRPT, which is often fair or limited in
its unfairness (see the discussion in \cite{wierman2011fairness}), 
the proposed variations using prediction have very poor fairness. 

This occurs for multiple reasons.  Most significantly, even very short
jobs can be caught waiting for jobs with a remaining predicted service
of zero.  This leads to occasional large values of $T(x)/x$ for small
jobs, skewing the fairness.  Also, in cases where small jobs obtain
large predictions, the value of $T(x)/x$ can again be high under high
load.   

The two problems suggest different solutions. In the first case,
$T(x)$ is artificially high; this suggests strong fairness results
require policies that do something beyond letting jobs with predicted
remaining service time zero continue.  Some addition of processor
sharing, preemption, or modifying the prediction when the predicted
remaining service time reaches zero should therefore improve fairness.
In the second case, $x$ is small when the predicted time $y$ is large;
this suggests that \added{the responsibility of a given job's slowdown is shared between the scheduler and the component responsible for service time prediction. To more accurately illustrate the interplay between those components, we speculate that} alternative definitions of fairness, based on the
prediction $y$ as well as the actual service time $x$, should be
considered.

We provide some preliminary results on fairness in \cref{app:fairness}, and leave additional study 
for further work.

\eat{

M2 Q1Q4Q7Q5 M5 Q1Q4Q7Q5 P2 S1

1.26607 1.26068 1.19887 1.19985
1.40785 1.39255 1.3014 1.30151
1.61483 1.57381 1.44632 1.44329
1.94808 1.8413 1.66831 1.65765
2.61687 2.30862 2.07917 2.0507
3.39307 2.76993 2.50992 2.45936
4.5376 3.35087 3.08643 3.00383
5.48352 3.76252 3.50939 3.40559

M5 S1
1.28405 1.27333 1.25635 1.24718
1.42101 1.39645 1.36967 1.35538
1.63025 1.57631 1.53535 1.51446
1.98536 1.86148 1.80359 1.77068
2.74312 2.40665 2.33448 2.27462
3.67031 2.98785 2.91791 2.82386
5.14659 3.76639 3.70459 3.55931
6.47324 4.33269 4.28406 4.10366

M2 S3

1.69503 1.6527 1.19558 1.2771
2.06167 1.95243 1.2947 1.40213
2.60239 2.34074 1.43332 1.56949
3.48065 2.87076 1.64271 1.81219
5.24874 3.69276 2.0205 2.2269
7.28798 4.4177 2.40514 2.63217
10.3106 5.24268 2.89786 3.12918
12.7782 5.78941 3.24021 3.47023

M5 S3

1.63672 1.58607 1.26128 1.42696
1.952 1.83918 1.37191 1.58643
2.42776 2.18632 1.52983 1.80069
3.21674 2.69141 1.78254 2.11028
4.84852 3.54957 2.27169 2.64999
6.83667 4.3658 2.811 3.19441
9.99687 5.37394 3.53727 3.90224
12.8336 6.08202 4.05134 4.40977

Q5 P2 S1 [M2 M4 M5 M8 M6]
1.19985 1.30952 1.24718 1.2397 1.43443
1.30151 1.45296 1.35538 1.3377 1.5963
1.44329 1.6694 1.51446 1.47429 1.82357
1.65765 2.03992 1.77068 1.68162 2.18267
2.0507 2.8955 2.27462 2.0603 2.90062
2.45936 4.13772 2.82386 2.4542 3.76796
3.00383 6.66568 3.55931 2.97329 5.16089
3.40559 9.20494 4.10366 3.35998 6.33088

Q5 P2 S3 [M2 M4 M5 M8 M6]
1.2771 1.61137 1.42696 1.37467 1.57732
1.40213 1.82375 1.58643 1.49395 1.75622
1.56949 2.11174 1.80069 1.64926 1.99762
1.81219 2.55179 2.11028 1.86956 2.35733
2.2269 3.43665 2.64999 2.24181 3.03528
2.63217 4.55814 3.19441 2.60048 3.79516
3.12918 6.45366 3.90224 3.04672 4.81496
3.47023 7.76928 4.40977 3.35053 5.37783
}

\begin{figure}[thp]
    \centering
    \includegraphics[width=\plotwidth]{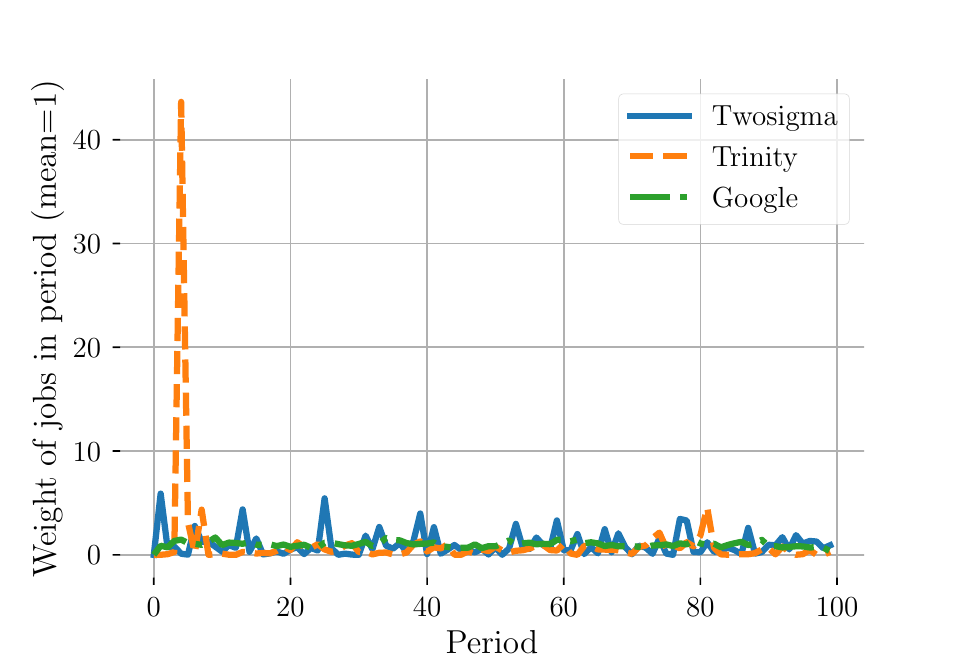}
    \caption{Cumulative weight of jobs submitted per period.}
    \label{fig:rw-weight-per-period}
\end{figure}

\begin{figure*}[thp]
    \centering
    \subfloat[Google]{\includegraphics[width=.32\textwidth]{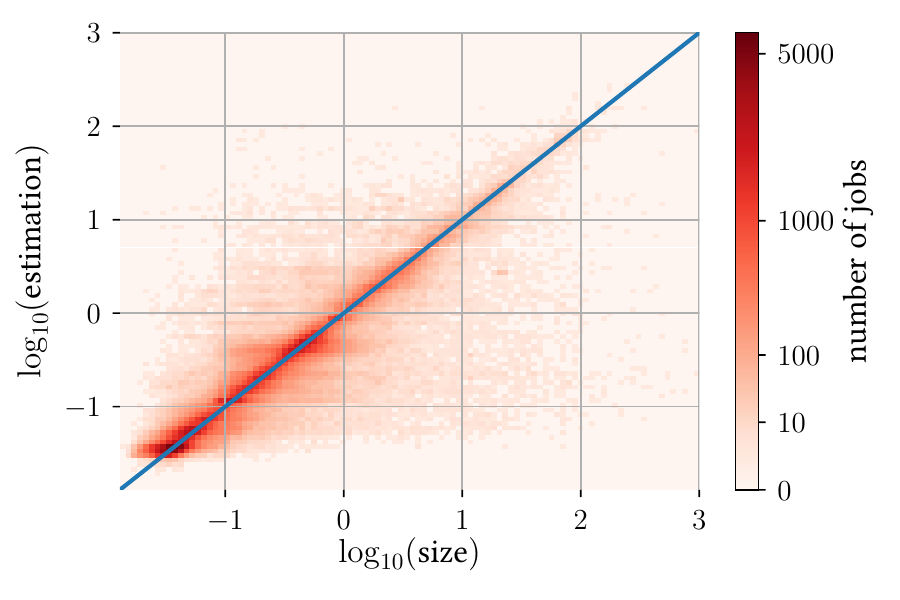}}
    \hfill
    \subfloat[Trinity]{\includegraphics[width=.32\textwidth]{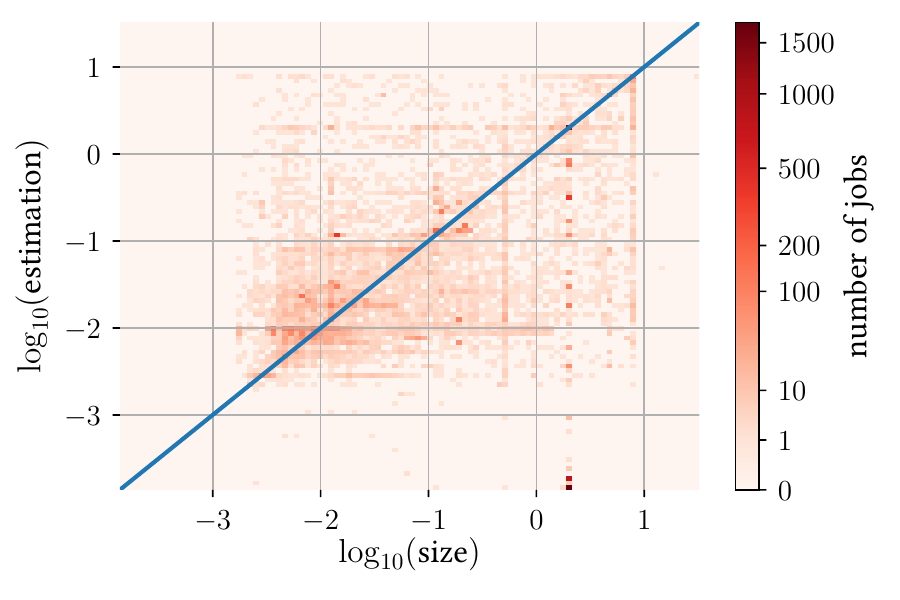}}
    \hfill
    \subfloat[Twosigma]{\includegraphics[width=.32\textwidth]{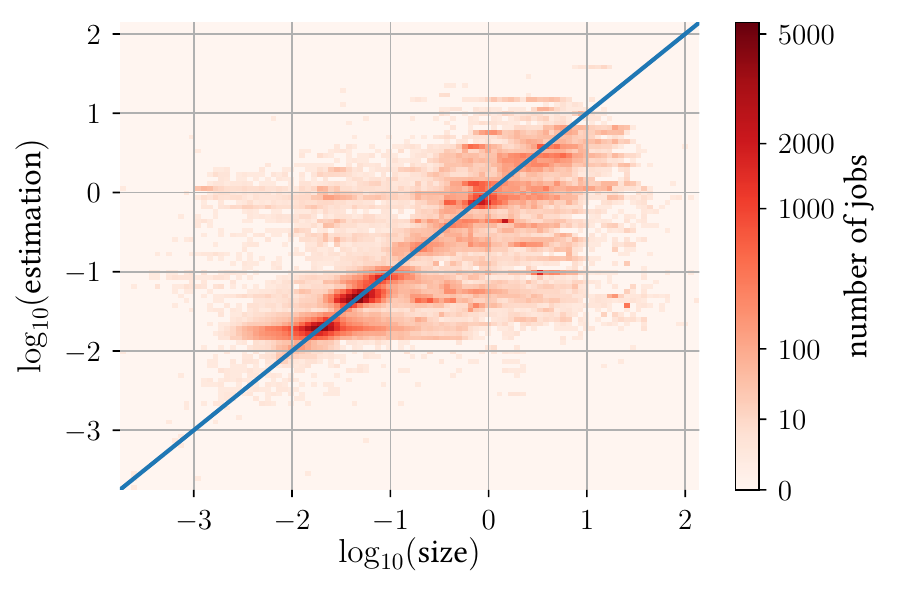}}
    \caption{Real-world datasets: heatmaps of \replaced{service time}{job size} distribution versus estimation.}
    \label{fig:rw-heatmaps}
\end{figure*}

\section{Real-World Traces}
\label{sec:realworld}


We now consider how the scheduling policies deal with real-world traces by Amvrosiadis et al.~\cite{amvrosiadis2018diversity}, who developed a system that predicts \replaced{service time}{job size} in large real-world clusters. The \replaced{service time}{job size} predictions are generated through a machine-learning system that focuses on detecting jobs that repeat on large compute clusters, using features such as user ID and job names in addition to information such as input size; further details are given by Tumanov et al.~\cite{tumanov2016jamaisvu}.
For each job, we have the submission time and its real and predicted runtime in seconds (for multi-task jobs, their \replaced{service time}{size}s were obtained by summing the \replaced{service time}{size} (resp. predicted \replaced{service time}{size}) of each individual task); we only take into account the jobs that completed successfully. For consistency with the synthetic traces, we normalize \replaced{service time}{job size} to obtain a mean of 1. We have three traces: Google (385,072 jobs), Trinity (18,872 jobs), and Twosigma (265,029 jobs). In \cref{fig:rw-weight-per-period}, we divide each trace in 100 periods, and plot the cumulative \replaced{service time}{size} of jobs submitted in that period, normalized such that the mean value per period is 1. We have very different job submission patterns: most of the load for the \replaced{Trinity}{Twosigma} dataset is concentrated around the beginning of the trace; the \replaced{Twosigma}{Trinity} dataset has lower load spikes along the whole trace; finally, the Google dataset has a relatively more uniform load pattern.\footnote{Amvrosiadis et al. discuss a fourth trace (Mustang). However, we have found that \replaced{this case falls in the extreme service time skew and inaccuracy scenario where predicted size-based scheduling is ineffective we referred to in \cref{sec:related}~\cite{dellamico2019scheduling,mitz2019scheduling}}{\replaced{service time}{job size} prediction in this case is not good enough to be useful for scheduling, and in our experiments scheduling policies that are not based on predicted \replaced{service time}{job size} outperform those that are}. For this reason, we do not include results from that dataset in this paper.}

To obtain an \emph{average system load} of $\lambda$ while maintaining the original patterns of job submission times, we normalize job submission times by multiplying them by a constant factor such that the average interval between jobs is $1/q\lambda$, where $q=100$ is the number of queues in the simulation. Since there is a degree of randomness due to the random choice of queues, we repeated the experiment 5 times for each settings, with negligible differences in the end results. Results shown here represent the average of the 5 runs.

\begin{figure*}[thp]
    \centering
    \subfloat[Shortest queue.]{\includegraphics[width=.32\textwidth]{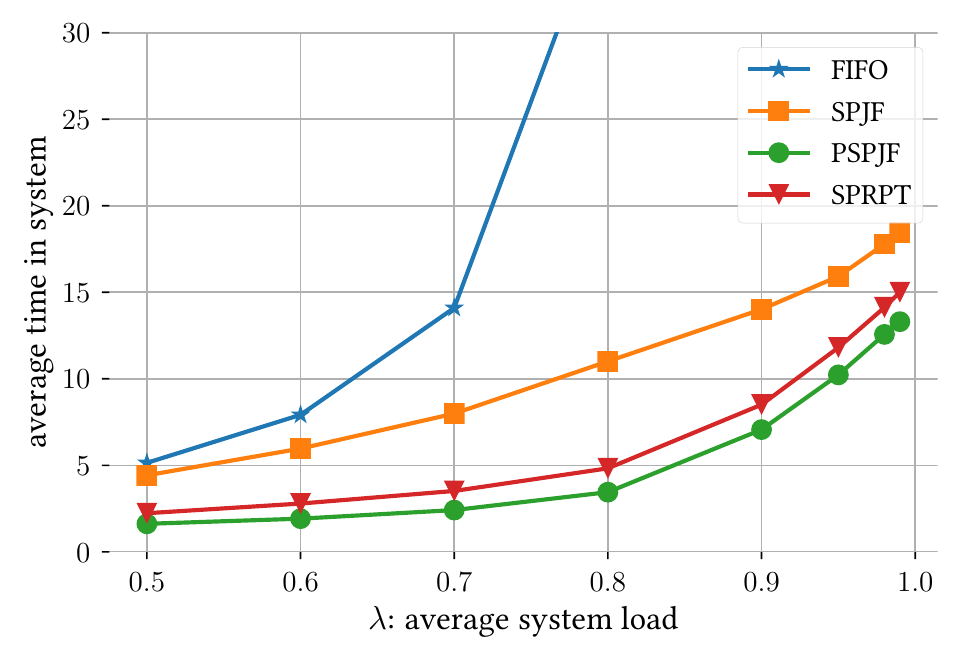}}
    \hfill
    \subfloat[Least loaded updated.]{\includegraphics[width=.32\textwidth]{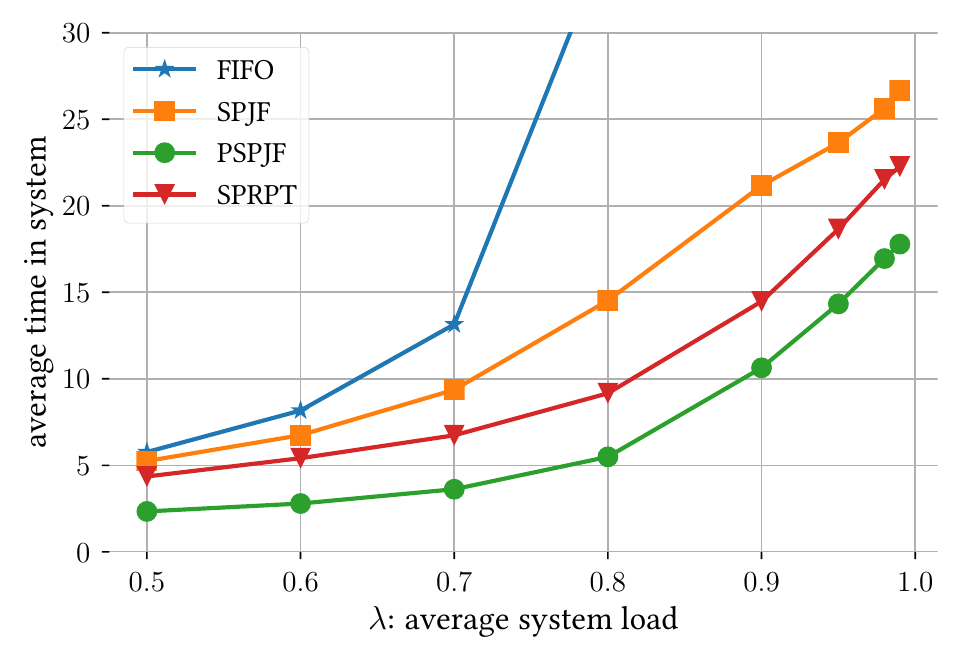}}
    \hfill
    \subfloat[Selfish queue selection.]{\includegraphics[width=.32\textwidth]{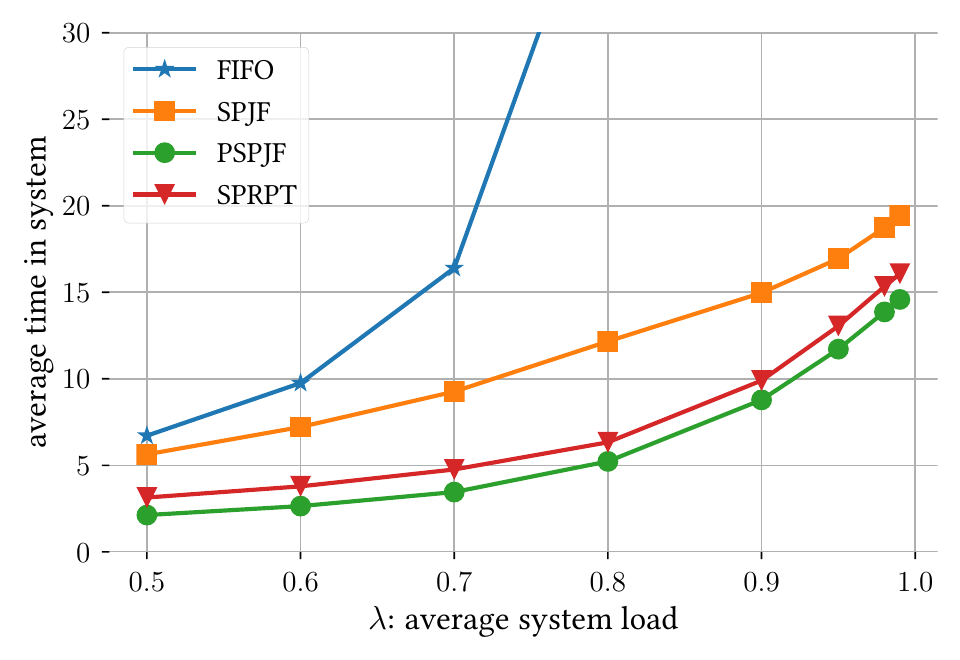}}
    \caption{Google dataset: mean \replaced{time in system}{response time} with various queue choice methods.}
    \label{fig:google-mst}
\end{figure*}

\begin{figure*}[thp]
    \centering
    \subfloat[Shortest queue.]{\includegraphics[width=.32\textwidth]{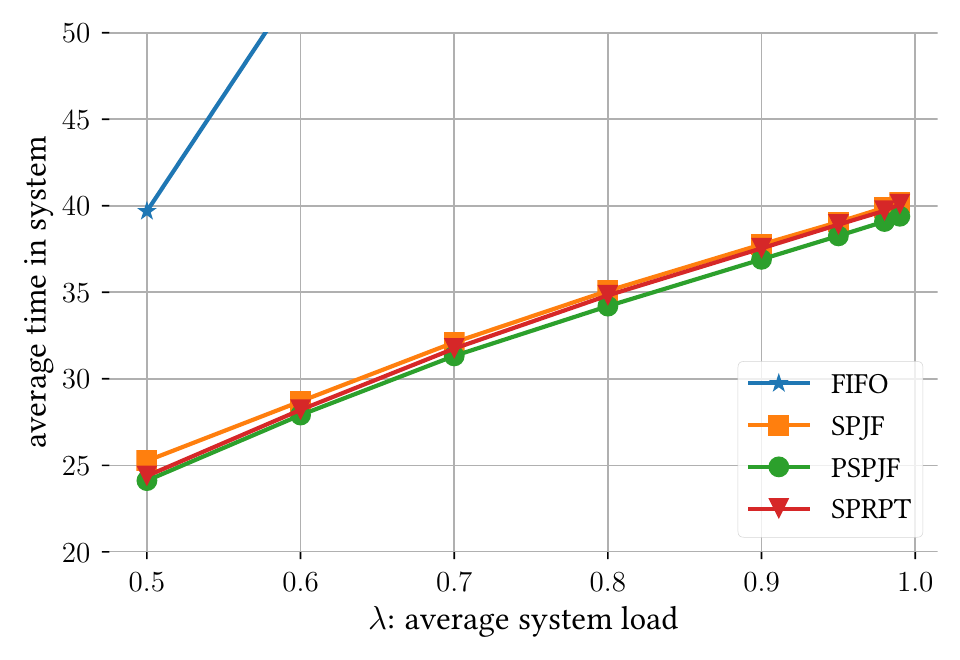}}
    \hfill
    \subfloat[Least loaded updated.]{\includegraphics[width=.32\textwidth]{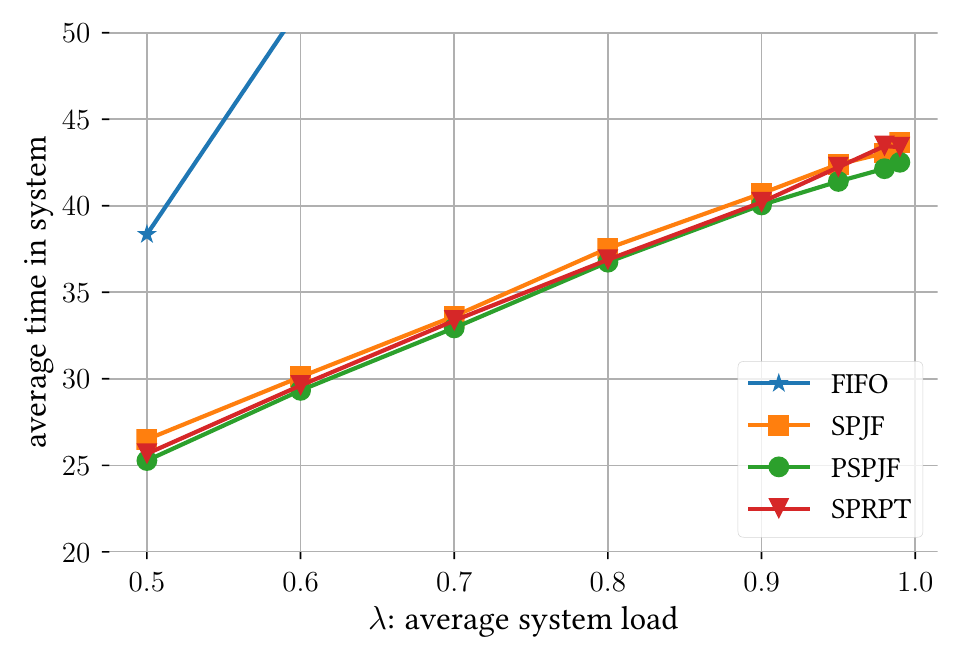}}
    \hfill
    \subfloat[Selfish queue selection.]{\includegraphics[width=.32\textwidth]{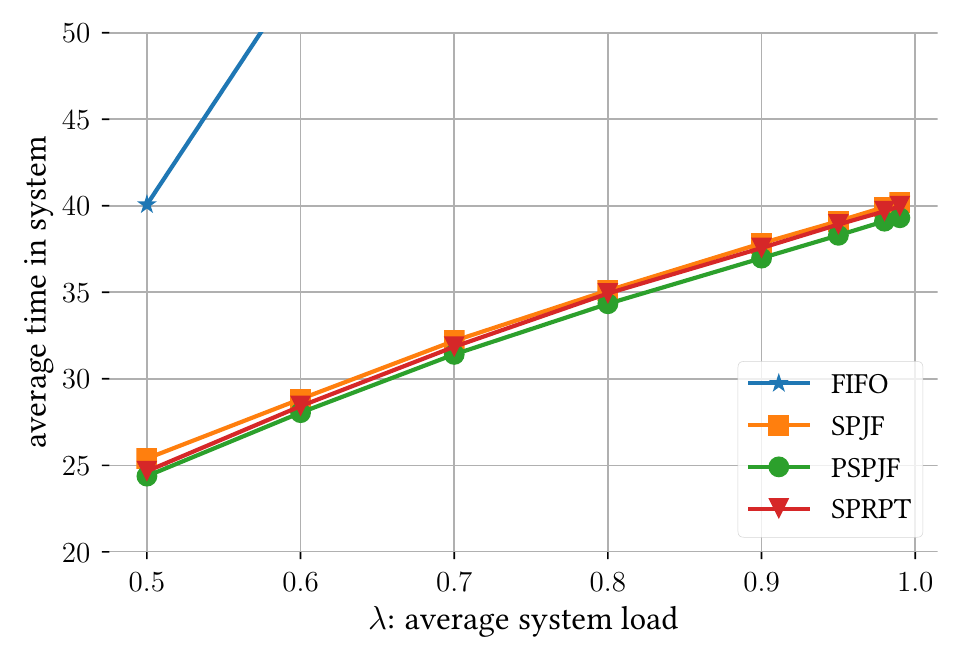}}
    \caption{Trinity dataset: mean \replaced{time in system}{response time} with various queue choice methods.}
    \label{fig:trinity-mst}
\end{figure*}

\begin{figure*}[thp]
    \centering
    \subfloat[Shortest queue.]{\includegraphics[width=.32\textwidth]{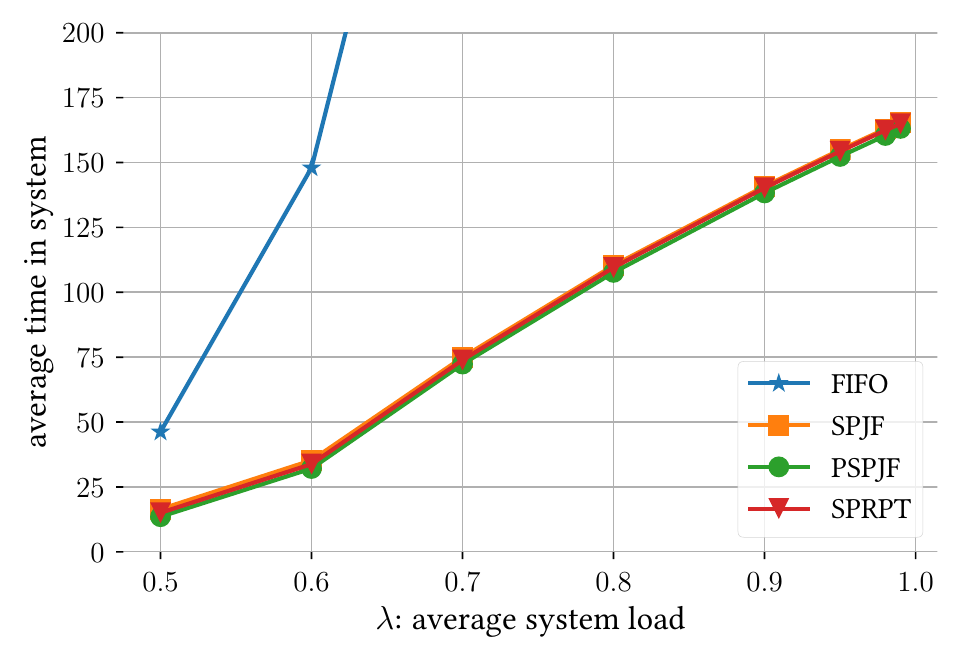}}
    \hfill
    \subfloat[Least loaded updated.]{\includegraphics[width=.32\textwidth]{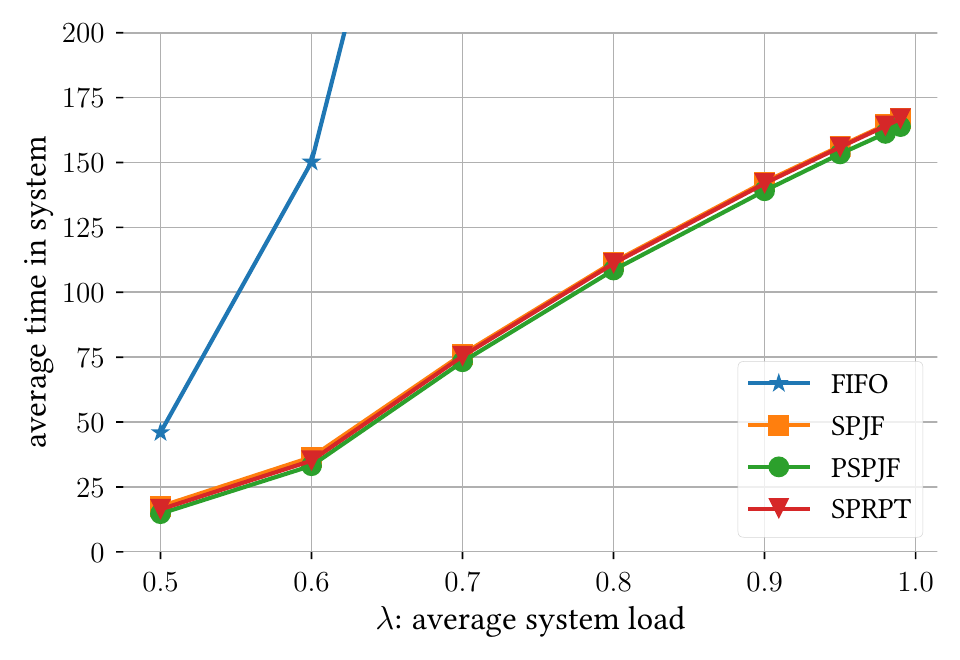}}
    \hfill
    \subfloat[Selfish queue selection.]{\includegraphics[width=.32\textwidth]{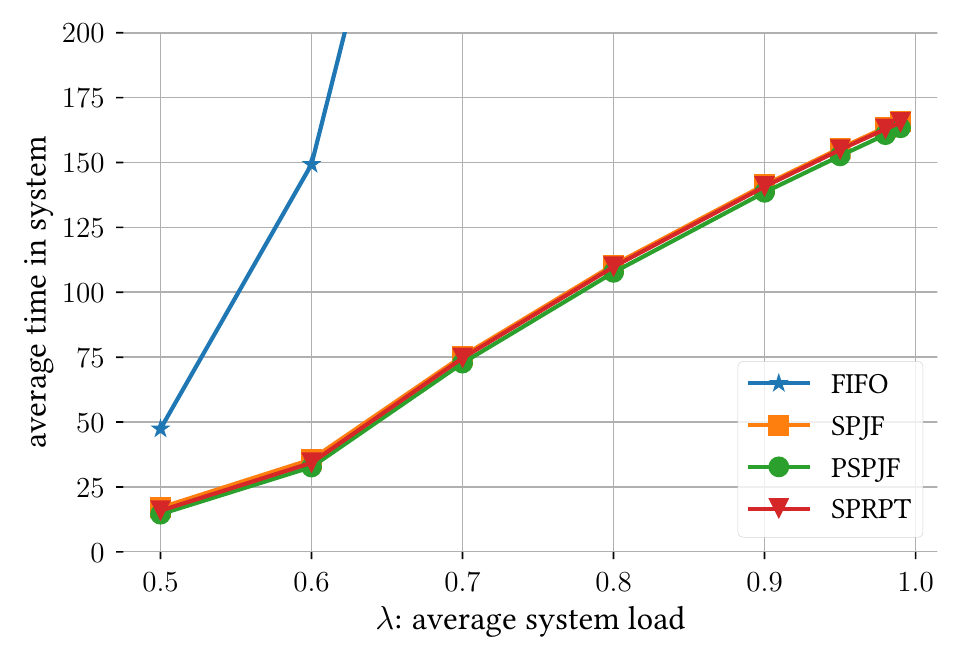}}
    \caption{\replaced{Twosigma}{Trinity} dataset: mean \replaced{time in system}{response time} with various queue choice methods.}
    \label{fig:twosigma-mst}
\end{figure*}

In \cref{fig:rw-heatmaps}, we show heatmaps covering the joint distribution of job real and predicted \replaced{service time}{size}. In what we believe is a common \replaced{real-world pattern}{pattern in real-world systems}, the \replaced{service time}{job size} distribution is heavy tailed, with a few large jobs representing a large amount of the overall system load; \replaced{service time}{job size}s are distributed along several orders of magnitude.


%
%

In \cref{fig:google-mst}, we show results for the Google dataset. The variation of system load in time yields higher absolute numbers for the mean \replaced{time in system}{response time} compared to the synthetic datasets seen before; this phenomenon is stronger in the other datasets because they have larger load spikes, as seen in \cref{fig:rw-weight-per-period}. Similarly to some use cases for synthetic workloads seen before, we observe that also here shortest queue performs better than least loaded updated. Interestingly, the selfish strategy instead performs essentially as well as shortest queue, showing that in this scenario the ``price of anarchy'' (i.e., the performance cost due to letting each job's owner selfishly choose their queue) is negligible\replaced{. Perhaps surprisingly, the selfish strategy outperforms least loaded updated. The main difference between the two strategies is that the former ignores the load due to larger jobs; as a result, and at the price of worse performance for larger jobs, smaller jobs get served earlier. Since smaller jobs are the majority, the average response time is better with selfish queue selection}{; analogous considerations hold for the other datasets we consider in this section}. \replaced{We further note that, u}{U}nsurprisingly, the size-based scheduling policies largely outperform FIFO scheduling in this heavy-tailed workload, and---again confirming results for synthetic workloads---PSPJF is preferable to the other policies due to its better performance \added{in the case that is most problematic for other policies, i.e.,} when large jobs are underestimated\added{~\cite{DCM,dellamico2019scheduling}}.



In the other two datasets (\cref{fig:trinity-mst,fig:twosigma-mst}) results are qualitatively similar, even though the larger load spikes further increase the average time spent in the system. The difference between FIFO and the other policies is even larger, while PSPJF remains the best-performing one. The difference between policies is smaller, compared to the large advantage due to choosing a policy based on predicted \replaced{service time}{job size}. \added{We explain this phenomenon with the fact that, in these traces with large load spikes, most of the waiting time is spent in very long queues during these spikes. In such cases, size-based policies behave in a largely similar way, giving small jobs priority over large ones; this explains the large gaps between them and FIFO. The difference between them, i.e., how preemption is implemented, plays a comparatively small role. Conversely, in the more balanced Google dataset, most jobs spend their waiting time in shorter queues; in this case preemption can impact waiting time more frequently.}




\section{Conclusion}

We have considered, primarily through simulation, the supermarket
model in the setting where service times are predicted.  As a starting
point, however, we considered the baseline where service times are
known.  Our results show that in the ``standard'' supermarket model
(exponential service times, Poisson arrivals) as well as more
generally, even though the \replaced{supermarket model}{power of two choices} provides tremendous
gains over a single choice, there remains substantial further
performance gains to be achieved when one makes use of known service
times.  In particular, \deleted{using} a service-aware scheduling policy such as
SRPT can yield significant performance gains under high loads.  This
\deleted{immediately} raises natural theoretical questions, such as deriving
equations for the supermarket model using least loaded server
selection with shortest job first or shortest remaining processing
time, which would extend the recent work of \cite{DBLP:journals/pomacs/HellemansH18}.  

However, our more important direction is to introduce the idea of
using {\em predicted service times} in this setting.  Our
simulation-based study allows us to evaluate this setting also on real-world traces, and suggests that the \replaced{supermarket model}{power of two choices}
maintains most of its power even when using predictions\added{, as long as they are reasonably precise}. We also find
some interesting, potentially counterintuitive effects.  \replaced{For example,
\replaced{choosing queue by length rather than predicted load can be beneficial when predictions are also used within the queue}{using queue lengths rather than the predicted load when choosing a queue can lead to better performance when also using the predictions to schedule within the queue}, because when predictions are sufficiently inaccurate using them for both choosing a queue and scheduling within \replaced{it}{the queue} can be detrimental.}{For example,
when predictions are sufficiently inaccurate performance is better when
using queue lengths rather than the predicted load when choosing a queue,
even when using the predictions to schedule within the queue.}
We view our results as showing the use of
predicted service times in large-scale distributed systems can be
quite promising in terms of improving performance.

\added{A practical piece of advice we can give for implementation in concrete systems is using \deleted{a} shortest queue selection and PSPJF\deleted{ per-queue scheduling}. This policy is simple to understand and implement, and our results show it performs very well in a quite wide variety of cases, being outperformed by non size-based scheduling only in extreme cases where predictions are particularly non-informative. While SPRPT for example can be a better scheduling policy when predictions are quite accurate, PSJPF appears more robust.  To evaluate the benefits of this policy and to possibly find better solutions for particular use cases, a further step can be replaying system traces in a simulator as done in this work.}

Our work highlights many open practical questions on how to optimize
these kinds of systems when using predictions, as well as many open
theoretical questions regarding how to analyze these kinds of systems.
For example, suitable mechanisms for managing jobs that exceed their
predicted service time offer further potential for important
improvements.  Perhaps most interesting is developing appropriate 
theories of fairness when using predictions.  A short job predicted
to have a long service time may face long delays before service;
how to achieve a suitable notion of fairness when using predictions 
clearly merits further study.

\section*{Acknowledgment}

The authors would like to thank Amvrosiadis et al.~\cite{amvrosiadis2018diversity} for sharing their
predictions datasets.

\balance

\begin{IEEEbiography}
        [{\includegraphics[width=1in,height=1.25in,clip,keepaspectratio]{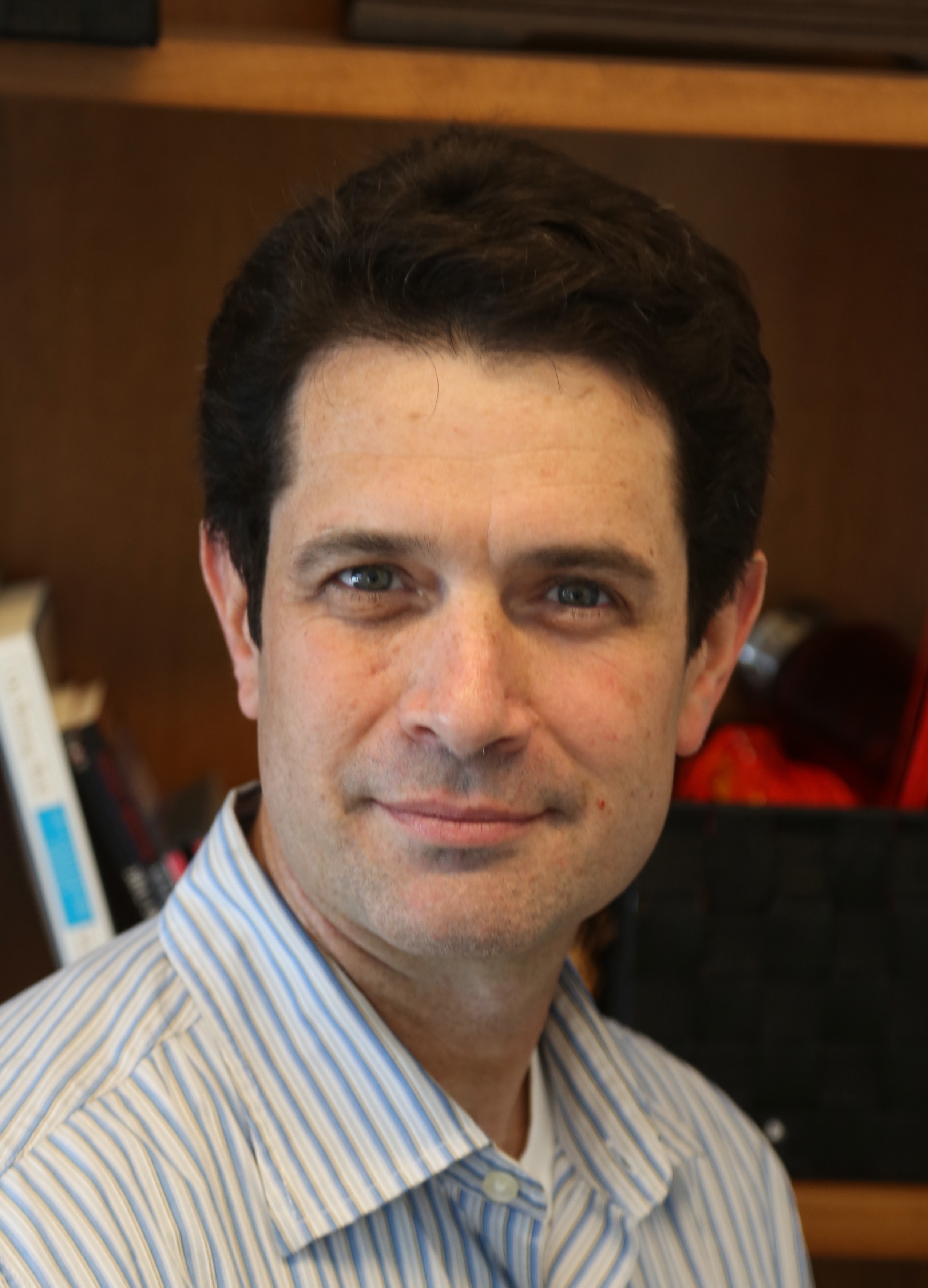}}]
        {Michael Mitzenmacher}
is a Professor of Computer Science in the School of Engineering and Applied Sciences at Harvard University. 
He works generally on algorithms and data structures for information and communication.   He has co-authored a textbook on randomized algorithms and probabilistic techniques in computer science published by Cambridge University Press.  He is an ACM and IEEE Fellow.
\end{IEEEbiography}

\begin{IEEEbiography}
        [{\includegraphics[width=1in,height=1.25in,clip,keepaspectratio]{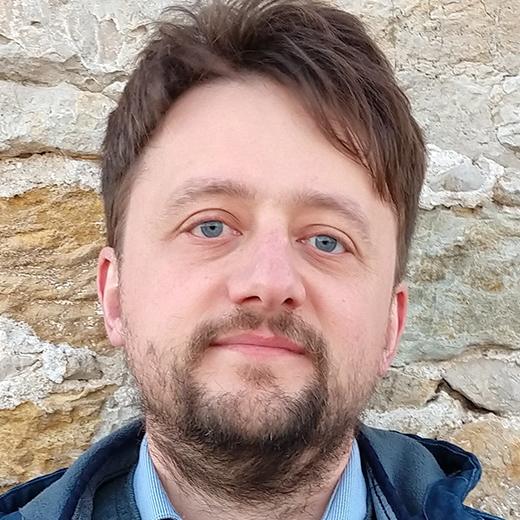}}]
        {Matteo Dell'Amico}
is an Assistant Professor of Computer Science at the University of Genoa. He works on distributed systems, computer security and data analytics.
\end{IEEEbiography}


%

\newpage


\appendices

\section{Additional Results}

\subsection{\texorpdfstring{\deleted{Toward}}{} Developing Equations for Limiting Behavior}
\label{app:cq1}

Previous work has shown that, in the limiting supermarket model
where the number of queues goes to infinity, individual queues can be
treated as independent, both when the choosing shortest queue and when
choosing the least loaded \cite{bramson2012asymptotic}.  
This connection plays a key role in the analysis of the supermarket 
model when choosing the least loaded queue with FIFO scheduling in
\cite{DBLP:journals/pomacs/HellemansH18}, which yields
the stationary distribution for the queue load for individual queues
in the limit as $n$ goes to infinity.    
(See also \cite{DBLP:journals/pomacs/HellemansBH19} on this issue;
this is sometimes referred to as cavity process analysis.)
We can
conclude that in equilibrium, at each queue considered in isolation,
the least loaded variant of the
supermarket model has a load-dependent arrival process, 
given by a Poisson process of rate $\lambda(x)$ when the queue has service load $x$.  (See, for example,
\cite{bekker2004queues,marie1980calculating} for more on queues with load-dependent
arrival processes; note here the arrival rate depends on the workload,
not the total number of jobs in the queue.)  The least loaded variant
of the supermarket model when  using other scheduling schemes, such as
SJF,  PSJF,  and SRPT,  would similarly have  the  same load-dependent  arrival
process, as in equilibrium the workload distribution would be the same
regardless of the scheduling  scheme.  Hence we could develop formulae
for quantities such as the  expected \replaced{time in system}{response time} in equilibrium
in the supermarket model using the  least loaded queue and SRPT, {\em if} we
can  develop  an  analysis  of  a  single  queue  using  SRPT  with  a
load-dependent  arrival process  (and similarly  for  other scheduling
schemes).  We are not aware of any such analysis in the literature, \added{and we know of no current technique that provides such an analysis};
this is a natural and tantalizing open question.

We note that the supermarket model when jobs choose the shortest queue
also, as far as we know, has not been analyzed for SJF, PSJF, and
SRPT.  Here the arrival process at a queue in equilibrium can be given
by a Poisson process of rate $\lambda(n)$ when the queue has $n$ jobs
waiting.  Again, if we can develop an analysis of a single queue using
SRPT with a queue-length-dependent arrival process, we can use this to
analyze the supermarket model using SRPT (and similarly for other
scheduling schemes).  \added{Again, we are not aware of any such analysis in the literature, and this is a natural and tantalizing open question.}

%

\subsection{Choosing a Queue with Predictions}
\label{app:cq2}

As before, we consider methods for choosing a queue beyond the
queue with the (predicted) least load.  We consider
placing a job so that it minimizes the additional predicted waiting time,
based on the predicted waiting times for all jobs.  
Alternatively, if control is not centralized, we might consider {\em
selfish} jobs, that seek only to minimize {\em their own} predicted waiting time
when choosing a queue.

\begin{figure*}[htbp]
  \centering
  \fourfig
    {Exponential service times, $\alpha = 0.5$}{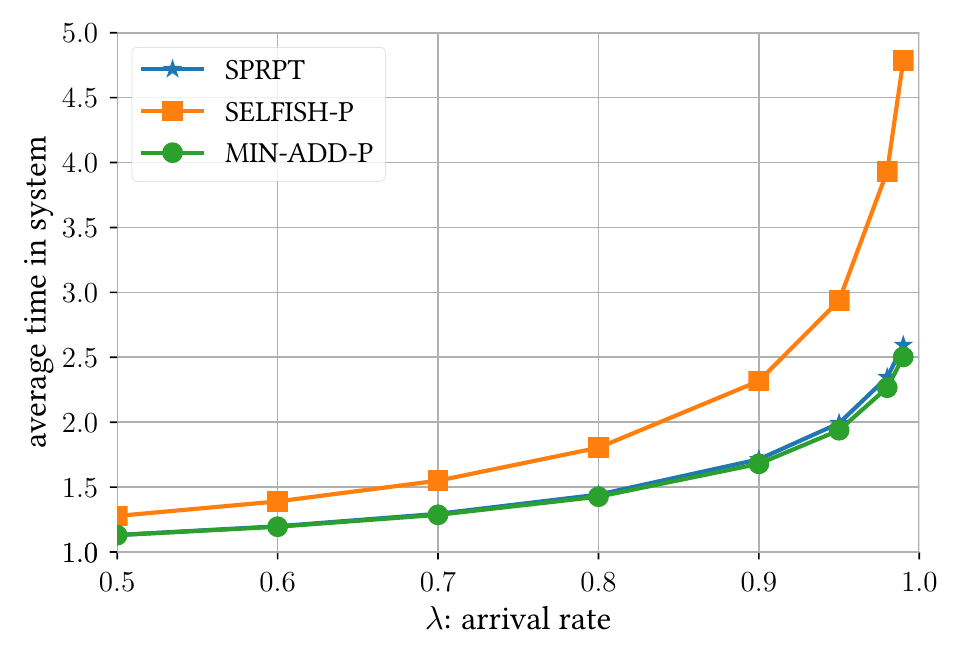}
    {Weibull service times, $\alpha = 0.5$}{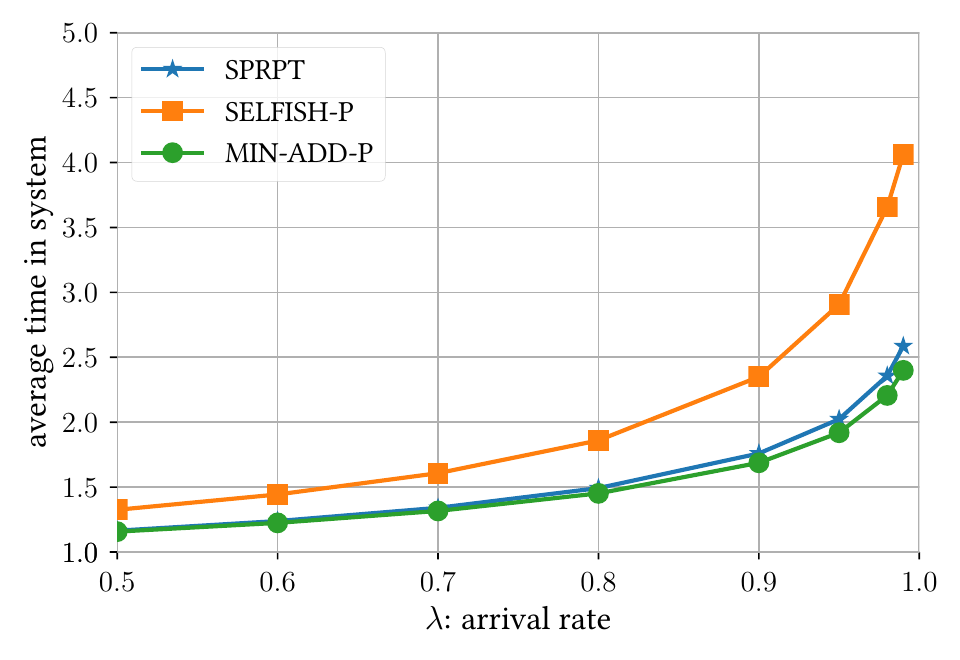}
    {Exponential service times, $\alpha = 0.5, \beta = 0.2$}{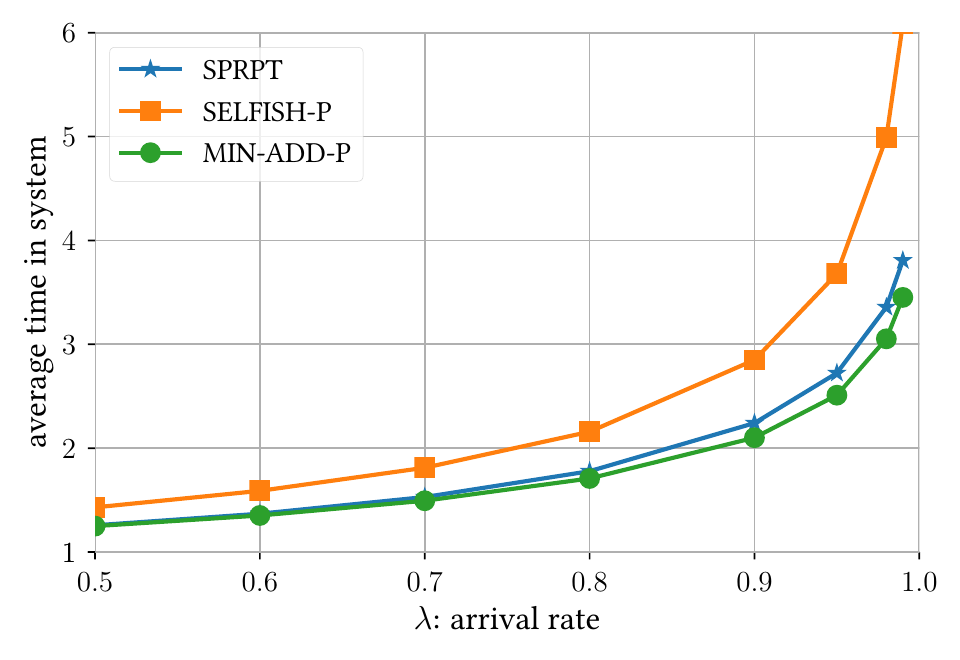}
    {Weibull service times, $\alpha = 0.5, \beta = 0.2$}{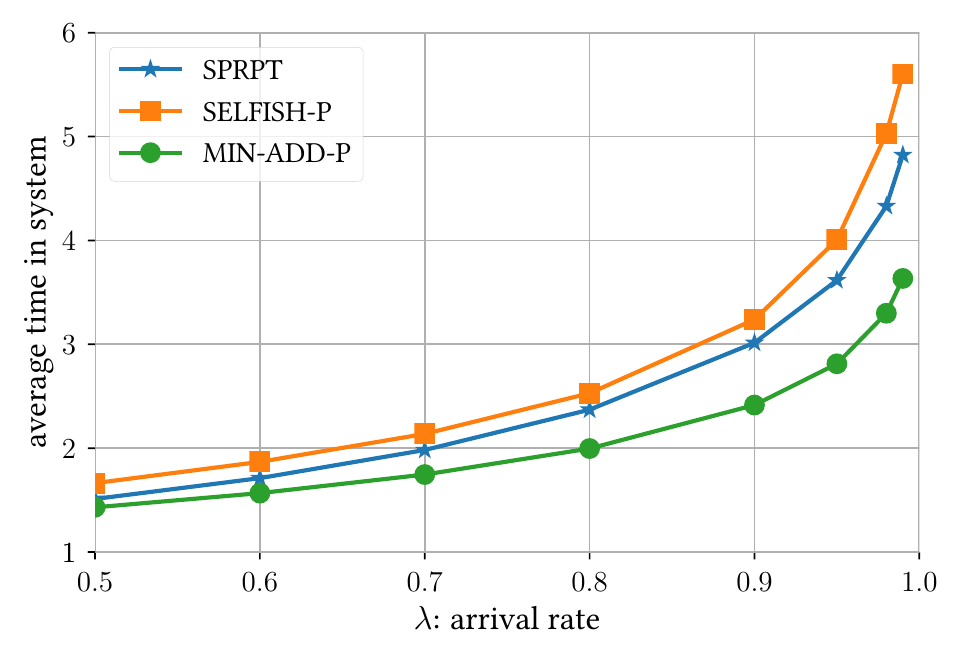}
\caption{Comparing methods of choosing a queue when using predicted service times, 
for $\alpha$-predictions with $\alpha = 0.5$, and for 
$(\alpha,\beta)$-predictions with $\alpha = 0.5, \beta = 0.2$.  All queues use SPRPT within the queue; in the figure, SPRPT means each job chooses the queue with smallest predicted remaining work, SELFISH-P means each job chooses
the queue that minimizes its own waiting time according to predictions, and MIN-ADD-P means each job chooses the queue that minimizes the 
additional waiting time added according to predictions.}
\label{fig:choosing2}
\end{figure*}

Our results, in Figure~\ref{fig:choosing2}, focus on two representative
examples: $\alpha$-predictions with $\alpha = 0.5$, and
$(\alpha,\beta)$-predictions with $\alpha = 0.5, \beta = 0.2$.  Again,
choosing a queue to minimize the additional predicted waiting time in
these situations does yield a small improvement over least loaded
update with SPRPT, and selfish jobs have a significant negative
effect.

\begin{figure*}[htp]
  \centering
  \twofig
    {Exponential service times, queue choice methods}{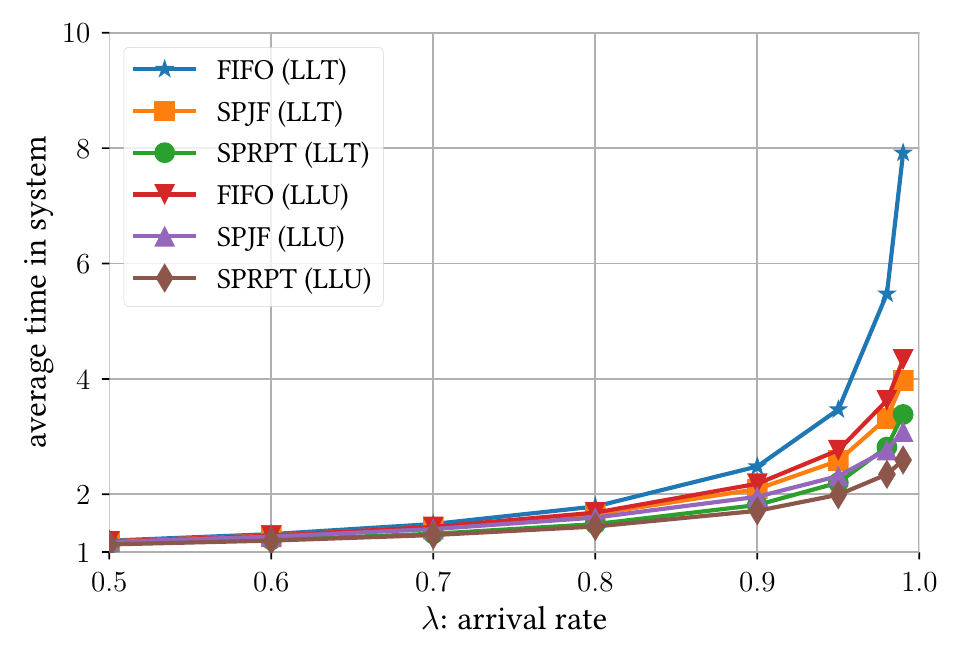}
    {Weibull service times, queue choice methods}{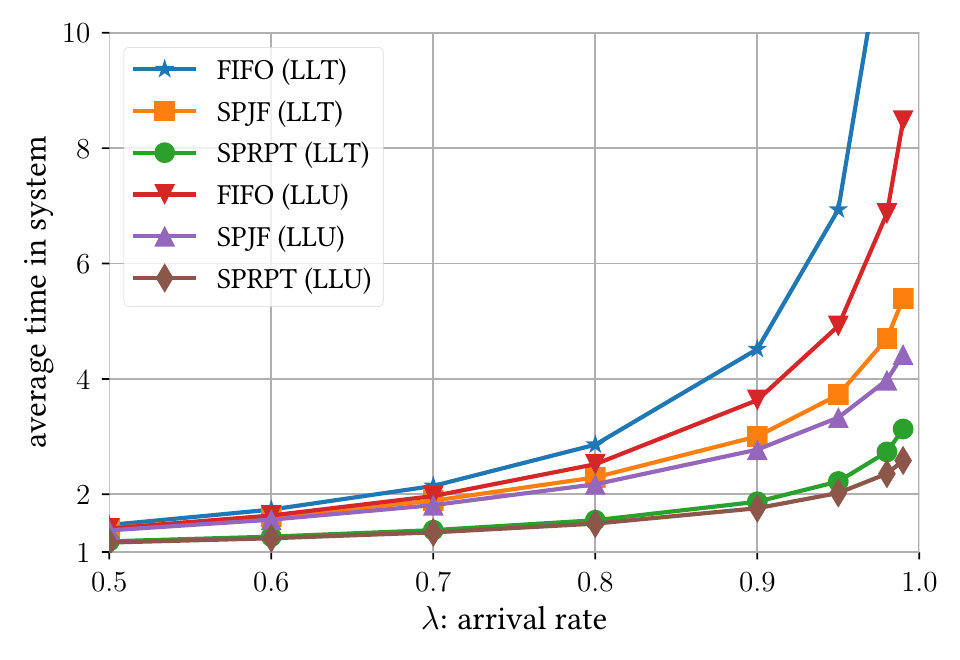}
\caption{Comparing variations of Least Loaded, for $\alpha$-predictions with $\alpha = 0.5$.}
\label{fig:llvlu}
\end{figure*}

Finally, as discussed earlier we note that there is a significant
difference between Least Loaded Updated and Least Loaded Total
policies.  Up to this point, we have used ``least loaded'' to refer to
Least Loaded Updated, where the predicted service time at the queue is
recomputed after each departure and arrival.  In contrast, Least
Loaded Total tracks a single predicted service time for the queue that
is updated on arrival but not at departure (unless a queue empties, in
which case the service is reset to 0).  While theoretically appealing
(as it reduces the state space for the system), Least Loaded Total
generally performs significantly worse than Least Loaded Updated.
Figure~\ref{fig:llvlu} below provides a representative example, in the setting
of $\alpha$-predictions when $\alpha = 0.5$.  We see FIFO, in particular,
does quite poorly under Least Loaded Total, and in all cases, the gap
in performance notably increases with the load.  Our other experiments show that
the gap in performance also increases significantly as the predictions
become more inaccurate; with exponential predictions, $\alpha$-predictions
with higher $\alpha$, or $(\alpha,\beta)$-predictions with $\alpha = 0.5$
and $\beta > 0$, our simulations show even larger gains from using Least Loaded
Updated.  While it may be useful to consider Least Loaded Total as an approach toward obtaining theoretical results, it does not appear to be the result we wish to aim for.

%
%
%


\section{Fairness}
\label{app:fairness}

As described in \cref{sec:fairness}, fairness is often defined in terms of making jobs spend an amount of time in the system that is roughly proportional to their \replaced{service time}{size}~\cite{wierman2011fairness}; this is captured by the \emph{slowdown} concept, where a job's slowdown is its resident time divided by its \replaced{service time}{size}. Here, we will show results relative to the real-world datasets of \cref{sec:realworld} where the average system load is set to $\lambda=0.9$.

\subsection{Slowdown Distribution}

\begin{figure*}
    \centering
    \subfloat[Shortest queue]{\includegraphics[width=.32\textwidth]{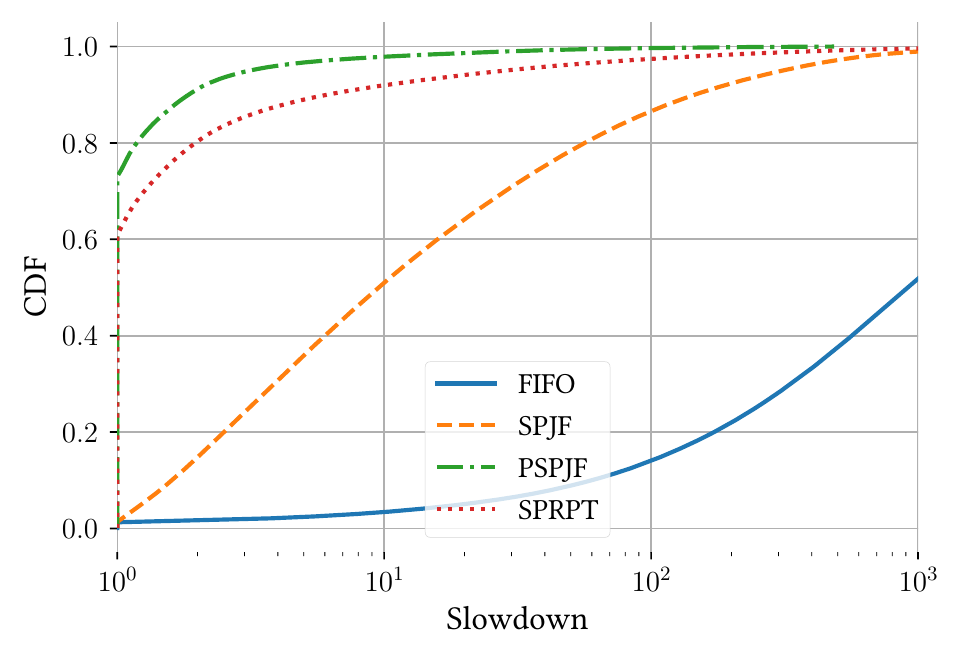}}
    \hfill
    \subfloat[Least loaded queue]{\includegraphics[width=.32\textwidth]{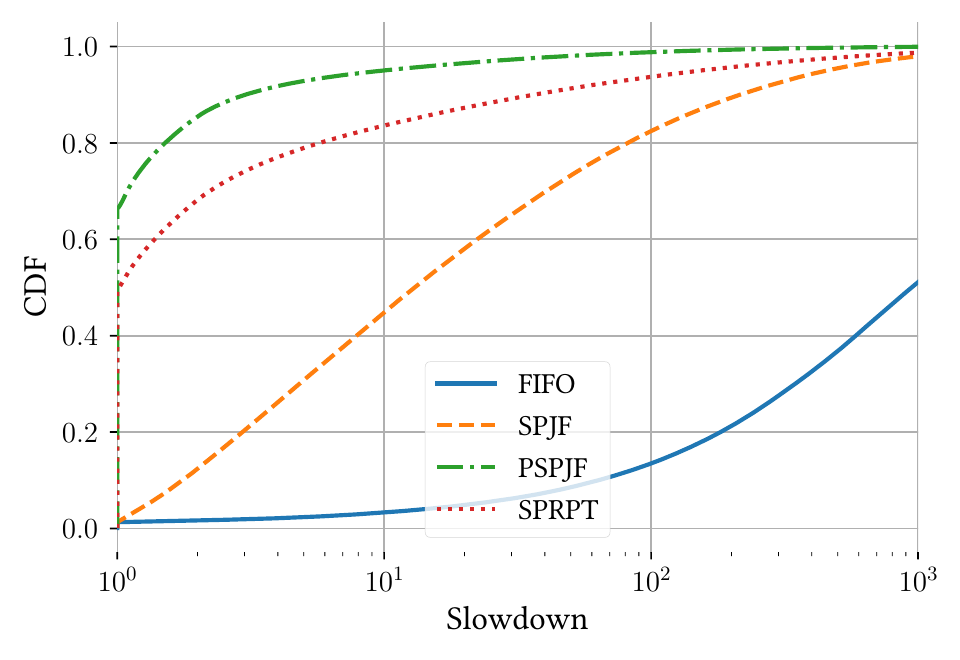}}
    \hfill
    \subfloat[Selfish queue selection]{\includegraphics[width=.32\textwidth]{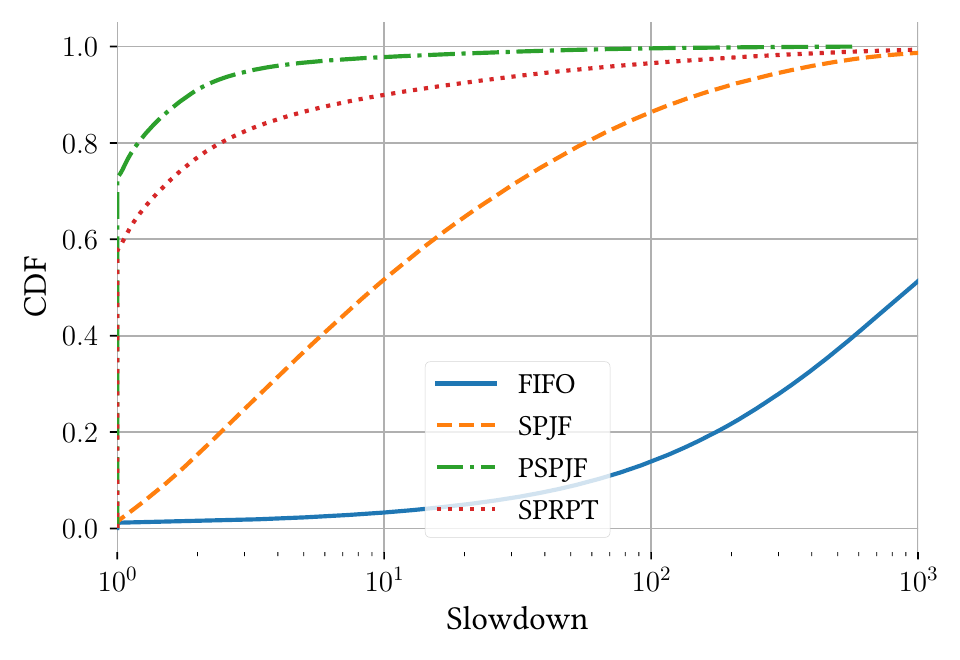}}
    \caption{Google dataset: CDF of slowdown.}
    \label{fig:google-slowdown}
\end{figure*}

\begin{figure*}
    \centering
    \subfloat[Shortest queue]{\includegraphics[width=.32\textwidth]{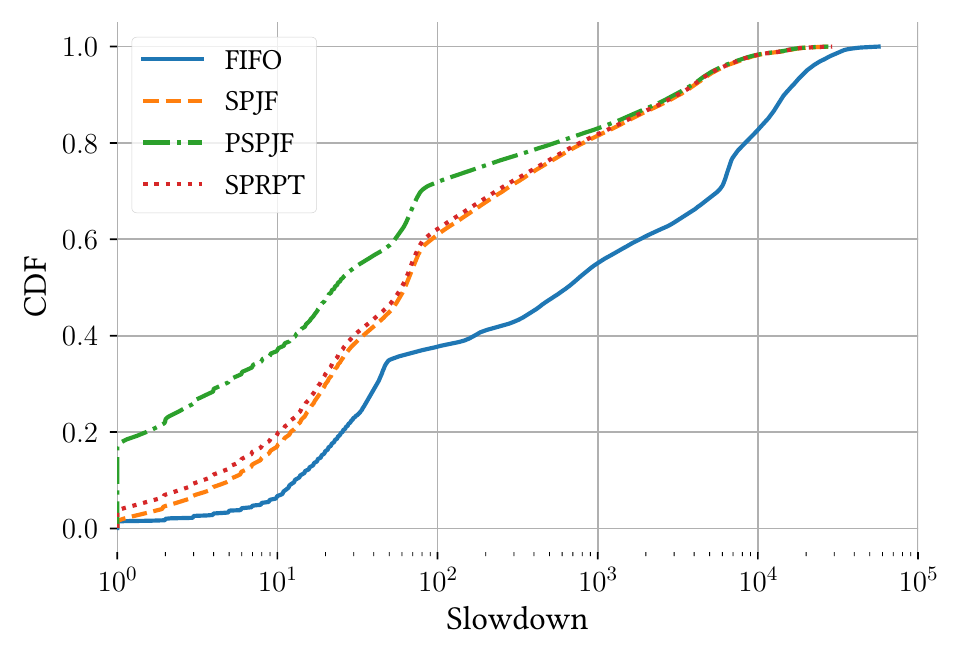}}
    \hfill
    \subfloat[Least loaded queue]{\includegraphics[width=.32\textwidth]{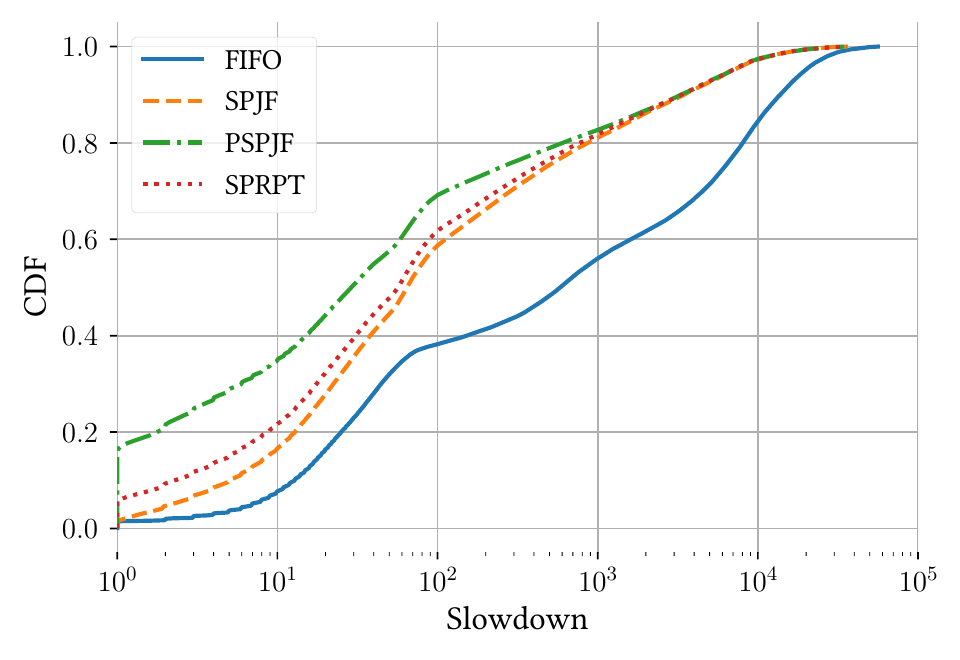}}
    \hfill
    \subfloat[Selfish queue selection]{\includegraphics[width=.32\textwidth]{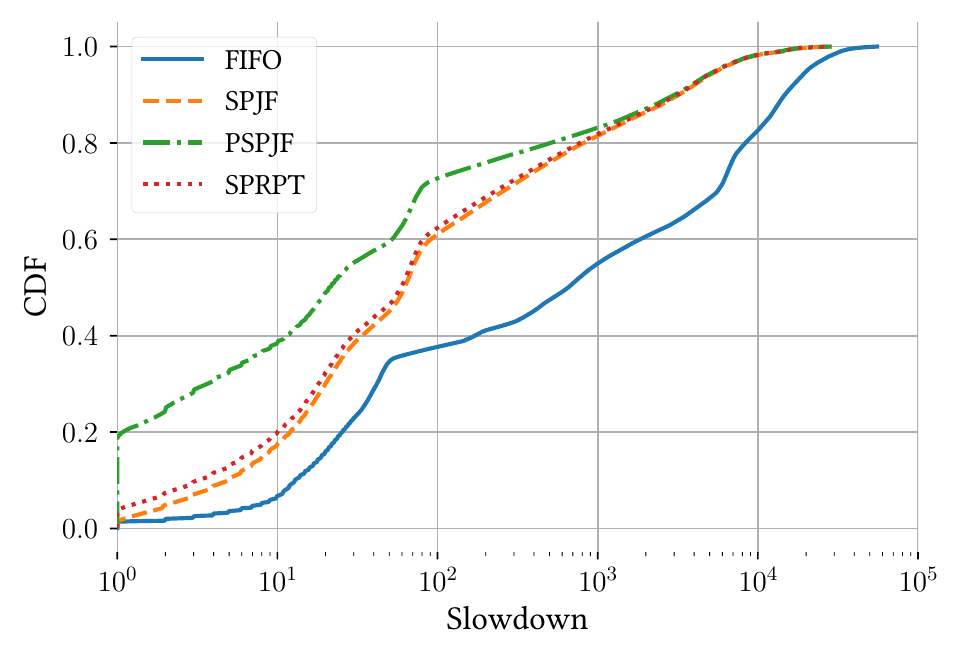}}
    \caption{Trinity dataset: CDF of slowdown.}
    \label{fig:trinity-slowdown}
\end{figure*}

\begin{figure*}
    \centering
    \subfloat[Shortest queue]{\includegraphics[width=.32\textwidth]{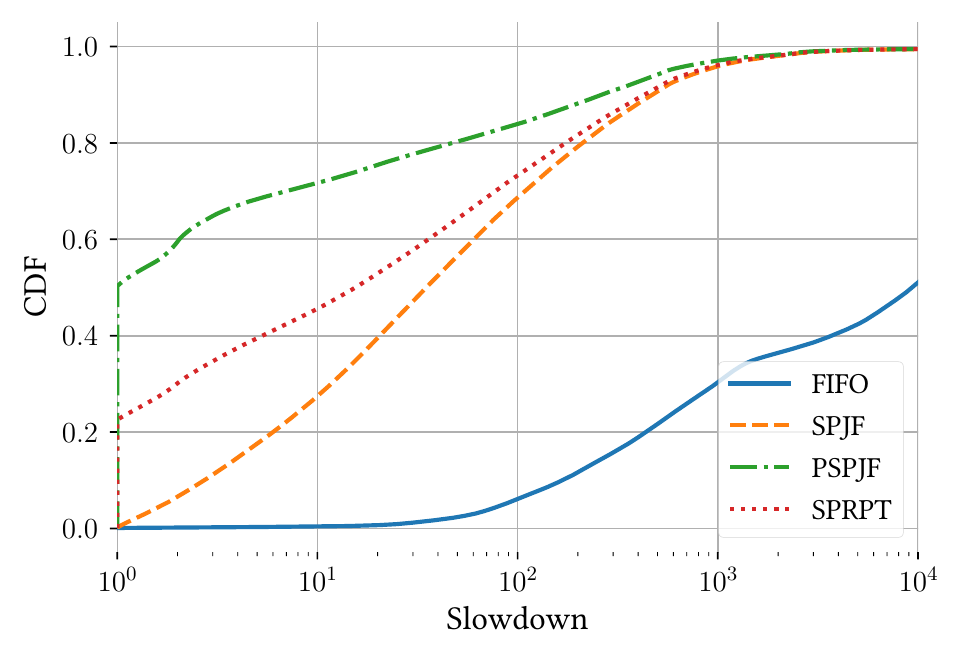}}
    \hfill
    \subfloat[Least loaded queue]{\includegraphics[width=.32\textwidth]{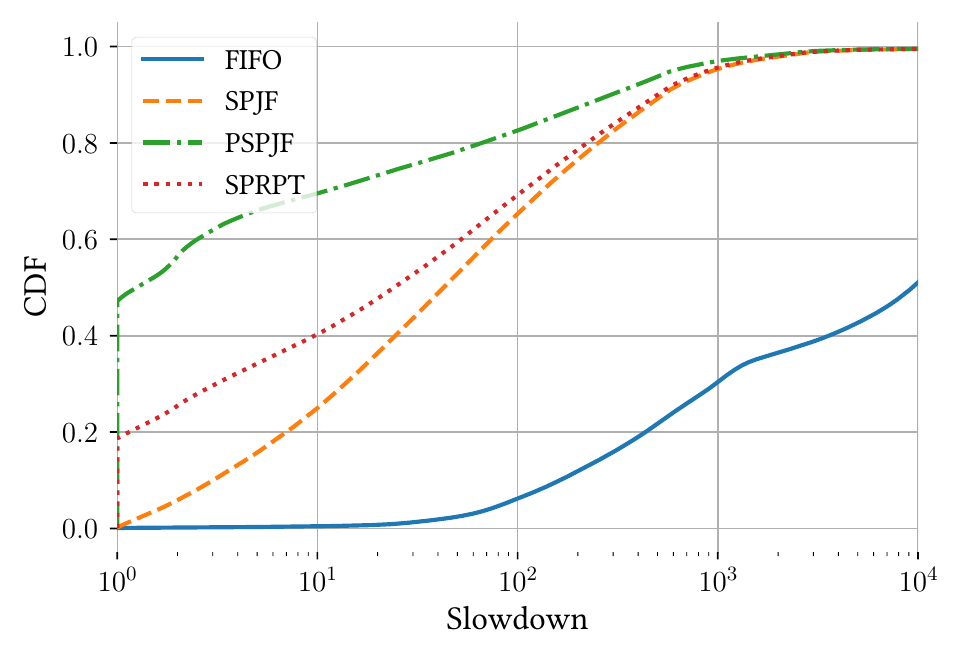}}
    \hfill
    \subfloat[Selfish queue selection]{\includegraphics[width=.32\textwidth]{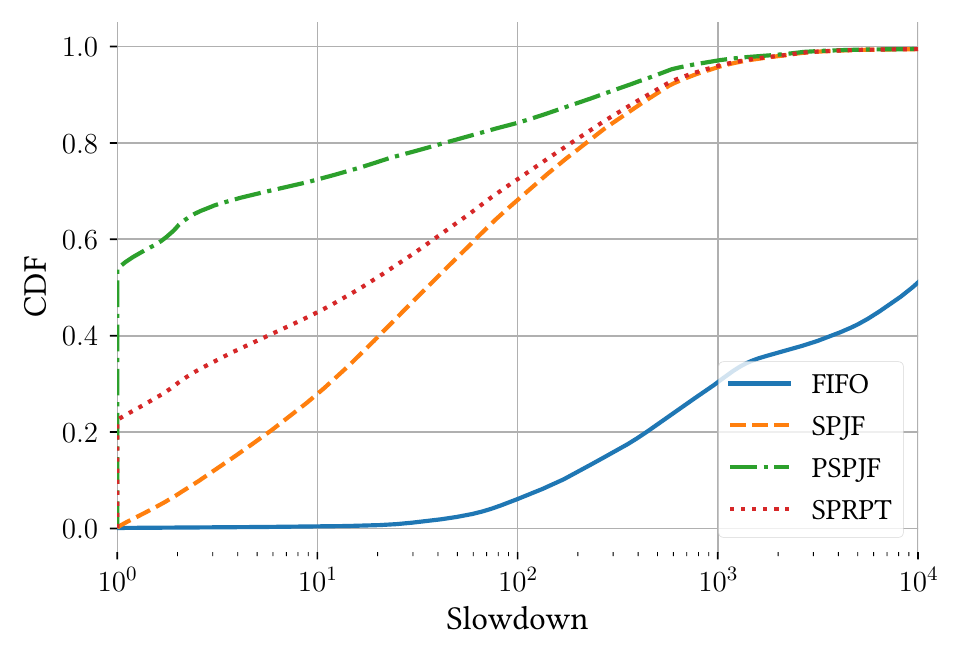}}
    \caption{Twosigma dataset: CDF of slowdown.}
    \label{fig:twosigma-slowdown}
\end{figure*}

A first definition of fairness can be having predictable slowdowns: for example, minimizing the variance of the per-job slowdowns, or a given value for the $x$-th percentile. To facilitate these evaluations, in \cref{fig:google-slowdown,fig:trinity-slowdown,fig:twosigma-slowdown} we show the empirical cumulative distribution functions (CDFs) of the slowdown values observed in our experiments.

Because system load changes over time, slowdown distributions are unequal in essentially all cases that we observe--in particular for Trinity, due to the large load the system experiences around the beginning of the trace. The overall pattern we see here, however, is that the slowdown distribution becomes less unequal with policies that perform better; PSPJF, which is the policy that performs best in terms of mean \replaced{time in system}{response time}, also has the least variability in terms of slowdown. We explain this intuitively with the fact that extreme slowdown values are caused by ``clogged'' queues: by optimizing mean \replaced{time in system}{response time}, extreme slowdown cases are also minimized.





\subsection{Mean Conditional Slowdown}

\begin{figure*}[htp]
    \centering
    \subfloat[Shortest queue]{\includegraphics[width=.32\textwidth]{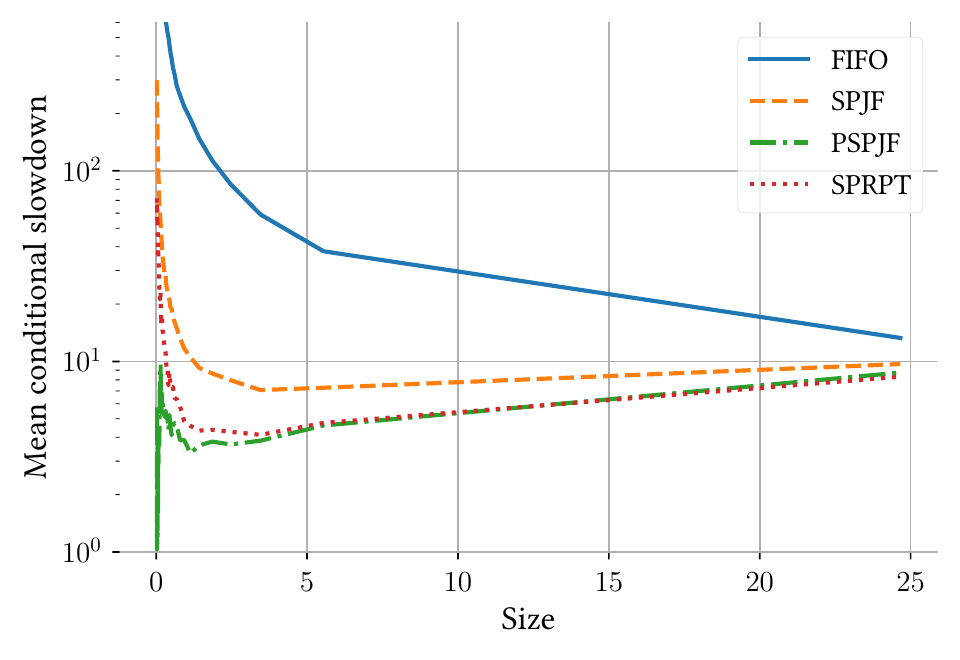}}
    \hfill
    \subfloat[Least loaded queue]{\includegraphics[width=.32\textwidth]{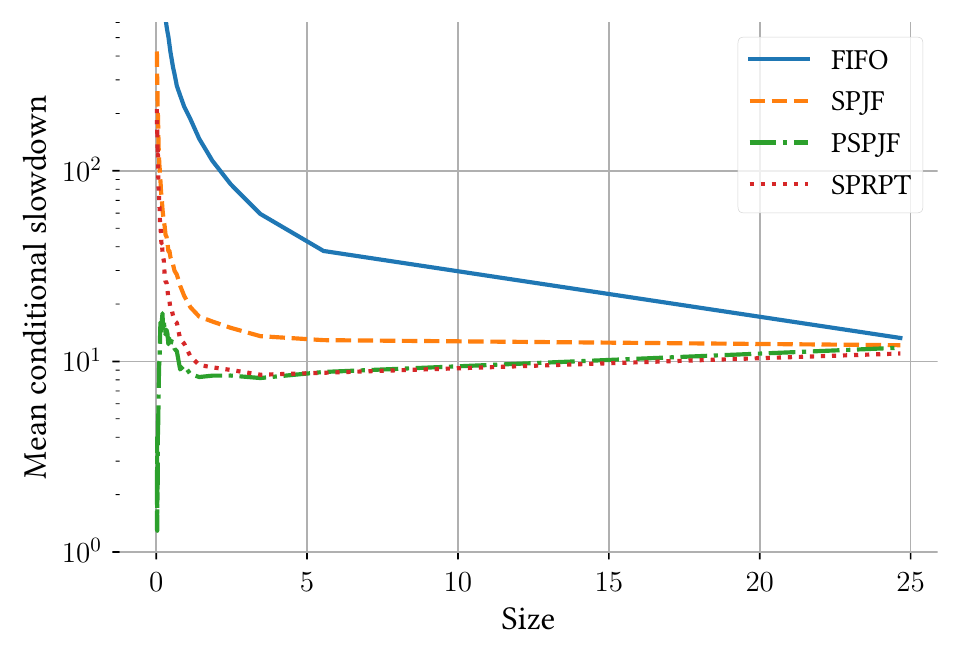}}
    \hfill
    \subfloat[Selfish queue selection]{\includegraphics[width=.32\textwidth]{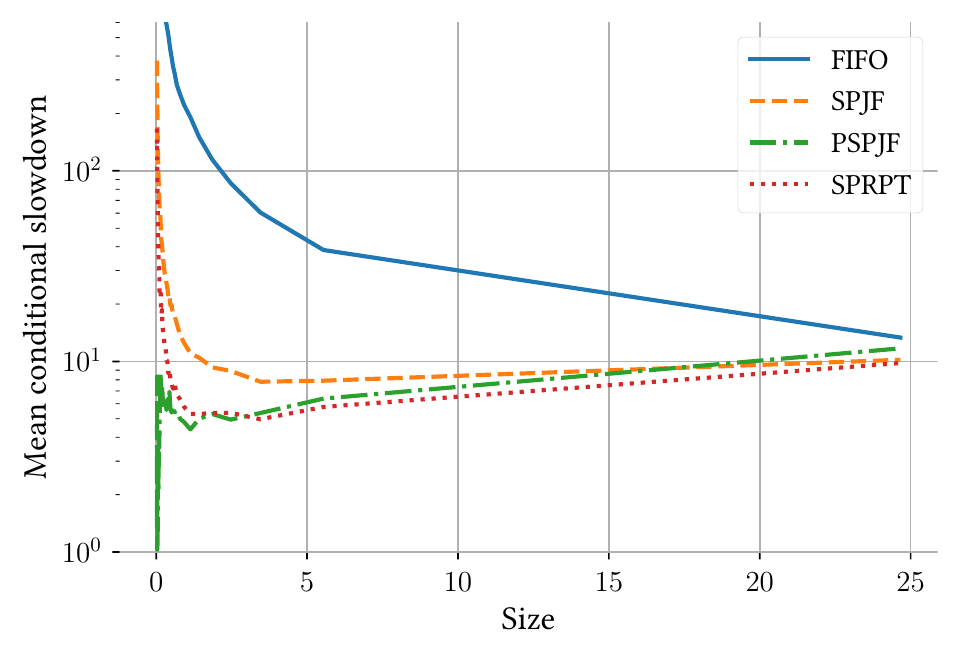}}
    \caption{Google dataset: mean conditional slowdown.}
    \label{fig:google-mcs}
\end{figure*}

\begin{figure*}[htp]
    \centering
    \subfloat[Shortest queue]{\includegraphics[width=.32\textwidth]{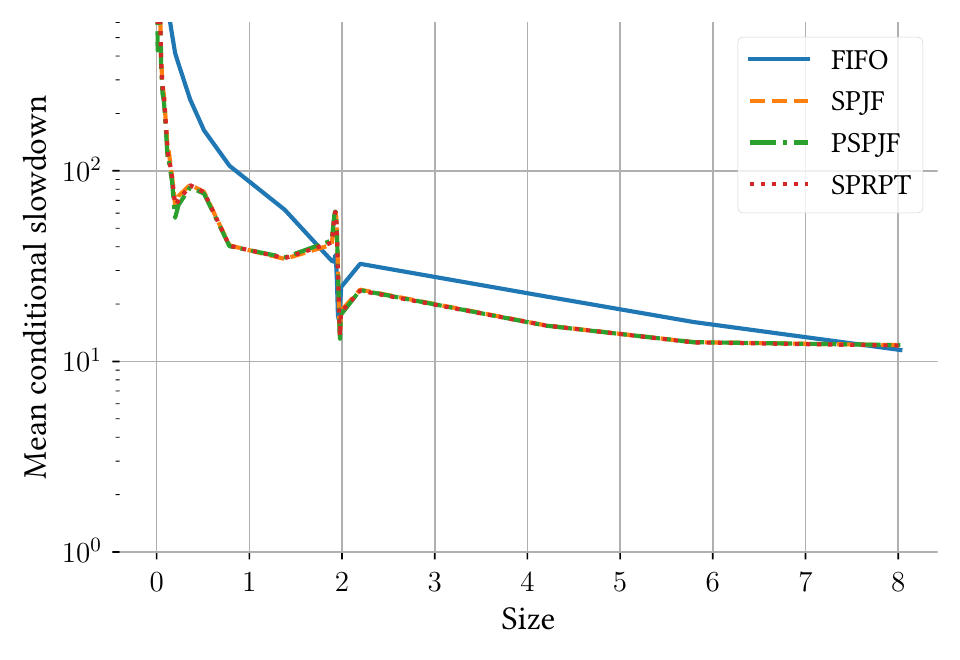}}
    \hfill
    \subfloat[Least loaded queue]{\includegraphics[width=.32\textwidth]{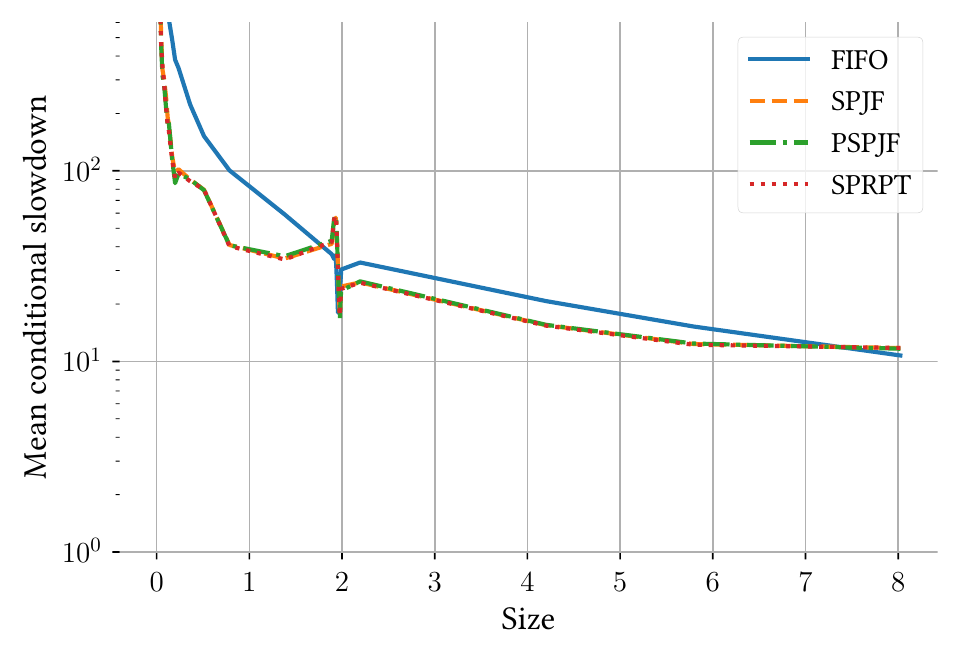}}
    \hfill
    \subfloat[Selfish queue selection]{\includegraphics[width=.32\textwidth]{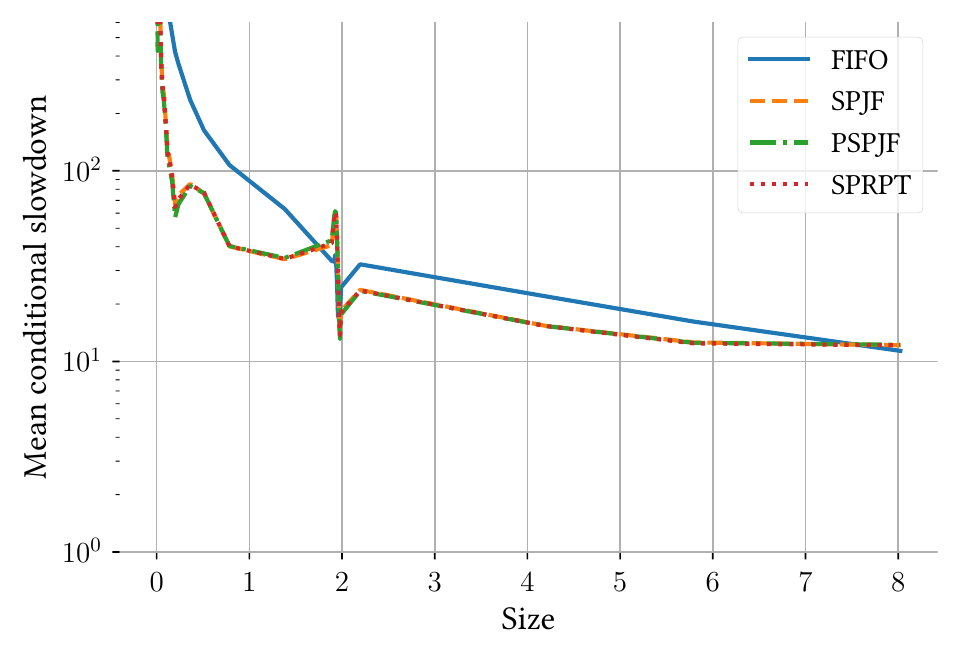}}
    \caption{Trinity dataset: mean conditional slowdown.}
    \label{fig:trinity-mcs}
\end{figure*}

\begin{figure*}[htp]
    \centering
    \subfloat[Shortest queue]{\includegraphics[width=.32\textwidth]{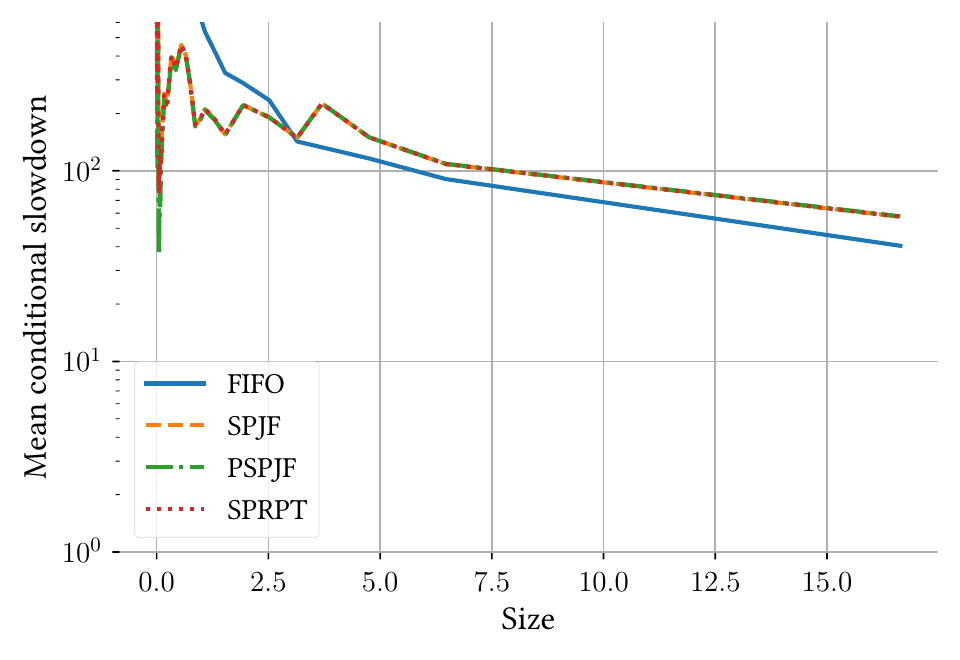}}
    \hfill
    \subfloat[Least loaded queue]{\includegraphics[width=.32\textwidth]{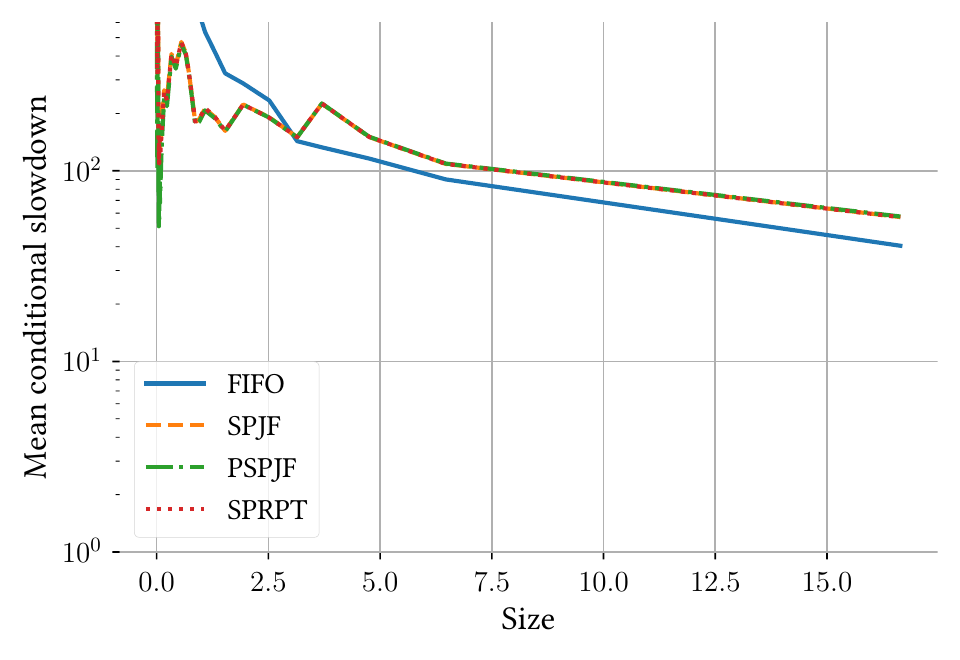}}
    \hfill
    \subfloat[Selfish queue selection]{\includegraphics[width=.32\textwidth]{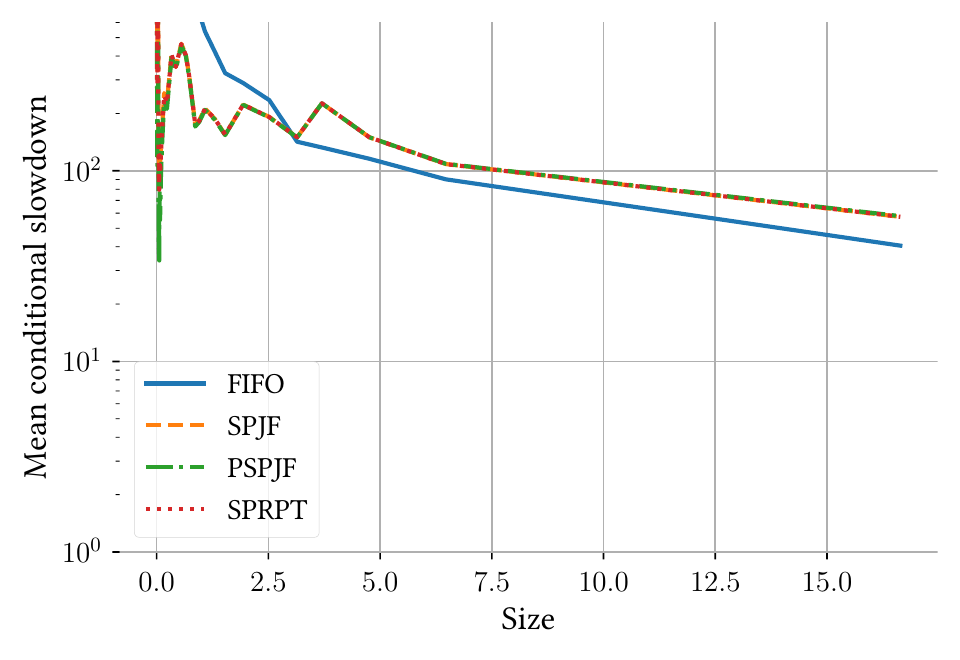}}
    \caption{Twosigma dataset: mean conditional slowdown.}
    \label{fig:twosigma-mcs}
\end{figure*}

\balance

A second definition of fairness involves \emph{mean conditional slowdown}, that is, the expected value of slowdown for job having a given \replaced{service time}{size}: for a job \replaced{having service time}{of size} $x$, the mean conditional slowdown is $E[T(x)]/x$, where $T(x)$ is the \replaced{time in system}{response time} for jobs \replaced{having service time}{of size} $x$ \cite{DBLP:conf/sigmetrics/BansalH01,wierman2011fairness,WiermanMHB}. To evaluate it empirically in our experiments, we follow the approach of Dell'Amico et al.~\cite{DCM}, bin jobs by \replaced{service time}{size} in 50 bins having the same amount of jobs each, and plot the average \replaced{service time}{size} and slowdown of each bin in \cref{fig:google-mcs,fig:trinity-mcs,fig:twosigma-mcs}, respectively on the X and Y axis.

Once again, and confirming esisting research~\cite{wierman2011fairness,DCM}, we observe that best-performing policies empirically result in better fairness. While one may imagine that size-based policies that give priority to smaller jobs would penalize large ones compared to a policy like FIFO, this is only true to some extent: for Google (\cref{fig:google-mcs}) size-based policies appeared to always be preferable also for larger jobs; in Trinity (\ref{fig:trinity-mcs}) only the very largest jobs are penalized; only in the Twosigma dataset large jobs have sensibly lower mean conditional slowdown when size-based policies are used.





In \cref{fig:trinity-mcs}, we notice a discontinuity for jobs of \replaced{service time}{size} 2. This can be explained by \cref{fig:rw-heatmaps}, which shows that for the Trinity dataset there is a set of jobs having \replaced{service time}{size} 2 whose \replaced{service time}{size} is systematically underestimated: this explains the drop in mean conditional slowdown.

For Twosigma (\cref{fig:twosigma-mcs}), the results for all the size-based scheduling algorithms look superimposed. We explain this, once again, with the data from \cref{fig:rw-heatmaps}, showing that \replaced{service time}{job size} estimations tend to be clustered around some well-separated values; hence, the details of the scheduling algorithm only have limited impact, since in most cases they will all schedule the job with the smallest estimated \replaced{service time}{size}.

\end{document}